\documentclass[proceedings]{ESI}
\usepackage{epsfig}
\usepackage{clpsref}

\title{Mesons as Bound States of Confined Quarks:\\
Zero and Finite Temperature}

\author{Pieter Maris\thanks{Present address: Dept.~of Physics,
North Carolina State University, Raleigh, NC 27695-8202, USA.} \ 
and Peter C. Tandy\\ 
Center for Nuclear Research, Department of Physics,\\
Kent State University, Kent OH 44242, USA.}

\conference{Confinement @ ESI, May-July 2000}

\abstract{
We survey recent work on the properties and decays of mesons as bound
states of confined quarks at both zero and finite temperature.  The
framework for these investigations is the set of QCD Dyson--Schwinger
equations truncated to ladder-rainbow level.  The infrared structure
of the ladder-rainbow kernel is described by two parameters; the
ultraviolet behavior is fixed by the one-loop renormalizaton group
behavior of QCD.  The work is restricted to the $u$, $d$ and $s$ quark
sector and allows a Poincar\'e-covariant study of the masses and
electroweak decay constants of the ground state pseudoscalars and
vectors: $\pi$, $K$, $\rho/\omega$, $K^\star$ and $\phi$.  Within the
impulse approximation, we summarize results for the $\pi$ and $K$
charge form factors.  Their timelike behavior exhibits the vector
meson production resonances and from this the associated vector meson
strong decay constants are extracted.  The confining property of the
quark self-energy dynamically produced in this work prevents the
appearance of spurious $\bar{q}q$ production effects.  The finite
temperature behavior of this model, and closely related ones, is
summarized for: the chiral restoration transition, the deconfinement
transition, the $T$-dependence of the spatial meson mode masses, and
the $T$-dependence of selected electroweak and strong decay widths.
The $T$-dependence obtained for meson masses is compared to results
from lattice QCD. }


\newcommand{\lsim}{\mathrel{\rlap{\lower4pt\hbox{\hskip0pt$\sim$}}
\raise1pt\hbox{$<$}}}
\newcommand{\gsim}{\mathrel{\rlap{\lower4pt\hbox{\hskip0pt$\sim$}}
\raise1pt\hbox{$>$}}}           

\newcommand{\case}[2]{\mbox{\small $\displaystyle \frac{#1}{#2}$}}
\newcommand{\Eq}[1]{Eq.~(\ref{#1})}

\newcommand{\Fig}[1]{Fig.~\ref{#1}}
\newcommand{\beq}{\begin{equation}}
\newcommand{\eeq}{\end{equation}}
\newcommand{\beqar}{\begin{eqnarray}}
\newcommand{\eeqar}{\end{eqnarray}}

\begin{document}

\hyphenation{ap-pro-xi-ma-ted}
\hyphenation{mo-men-tum}
\hyphenation{cross-over}
\hyphenation{de-pend-ence}
\hyphenation{in-de-pend-ent}
\hyphenation{smooth-ly}
\hyphenation{amp-lit-ude}
\hyphenation{pho-ton}
\hyphenation{Blas-chke}
\hyphenation{Schrod-inger}
\hyphenation{gro-und}
\hyphenation{mod-el}
\hyphenation{in-va-riant}
\hyphenation{rain-bow}
\hyphenation{electro-magn-etic}
\hyphenation{expon-ents}
\hyphenation{Nem-irov-sky}
\hyphenation{smooth-ly}
\hyphenation{res-torat-ion}
\hyphenation{trun-cat-ion}
\hyphenation{mod-el}
\hyphenation{mass-es}
\hyphenation{qu-arks}

\section{Introduction}
\label{secintro}
{\sloppy 
The light pseudoscalar mesons play an important role in understanding
low-energy QCD.  They are the lightest observable hadronic bound
states of a quark and an antiquark, and are the Goldstone bosons
associated with chiral symmetry breaking.  Their static properties
such as the mass and decay constants have been studied extensively and
are qualitatively understood within QCD.  A number of studies of
pseudoscalar mesons~\cite{mpifpi,fr,MR97} have played a key role in
the development of continuum methods for modeling QCD via the
Dyson--Schwinger equations [DSEs].  Such methods have now evolved 
into an excellent tool for the study of nonperturbative aspects of 
a variety hadronic properties and processes in
QCD~\cite{pctrev,bastirev,Alkofer:2000wg}.

Here we recall recent studies of light mesons, both at zero
temperature and at finite temperature, in which the quark propagator
DSE in the $u$, $d$ and $s$ quark sector is coupled to the meson bound
state Bethe--Salpeter equation [BSE].  The bound state amplitudes so
produced are used to study meson decays and form factors.  This
approach is consistent with quark and gluon
confinement~\cite{bastirev,Alkofer:2000wg,conf}, generates dynamical
chiral symmetry breaking~\cite{dcsb}, and is explicitly Poincar\'e
covariant.  It is straightforward to implement the correct one-loop
renormalization group behavior of QCD~\cite{MR97}, and to obtain
agreement with perturbation theory in the perturbative region.
Provided that the relevant Ward--Takahashi identities [WTI] are
preserved in the truncation of the DSEs, the corresponding currents
are conserved.  The axial WTI ensures the Goldstone nature of the
pions~\cite{MRT98}; the vector WTI is important for electromagnetic
current conservation.

We show results for the pion and kaon electromagnetic form factors in
impulse approximation~\cite{MT99pion,MTpiK00}, both in the timelike
and the spacelike region.  In impulse approximation, with quark
propagators dressed in rainbow approximation and meson Bethe--Salpeter
amplitudes produced in ladder approximation, the vector WTI for the
quark-photon vertex ensures that the meson electromagnetic current is
conserved.  A feature of this work is that the confining property of
the dynamically generated quark self-energy prevents the appearance of
spurious $\bar{q}q$ production effects in the timelike region.  The
ground state vector mesons provide the lowest physical resonances in
the structure of the timelike charge form factors of the pion and
kaon.  We summarize the ladder-rainbow DSE-BSE results for the ground
state vector mesons~\cite{MT99} as well as their strong and
electroweak decays.  The various strong and electromagnetic
interactions involving vector and pseudoscalar mesons have been a
typical testing ground for non-perturbative QCD models~\cite{pctrev}
and here we give a few examples of modern treatments of such
processes.

At finite temperature we discuss chiral restoration and deconfinement
using the Matsubara formalism.  In rainbow-ladder truncation, there is
a second-order, mean field chiral phase transition, with a critical
temperature around $120 \sim 150$ MeV depending on the details of the
ladder kernel~\cite{hmr98T}.  We also summarize results for the
temperature dependence of the spatial masses of the scalar,
pseudoscalar, and vector $\bar{q} q$ bound states and selected
electroweak and strong decay widths of these spatial $\bar{q} q$
modes~\cite{MRST00,sepT01}.  Despite deconfinement, these modes seem
to persist above the chiral and deconfinement phase transition.  The
results for the $T$-dependence of spatial meson masses are in
reasonable qualitative agreement with results from lattice QCD.

In the next Section we give the relevant DSEs, followed by a
discussion of the truncation and the model parameters in
Sec.~\ref{secladderrainbow}.  In Sec.~\ref{secff} we discuss in detail
the application of the DSE approach to calculate meson electromagnetic
form factors.  Modifications of the model to study QCD at finite
temperature are considered in Sec.~\ref{secextfinT} and there we also
discuss the chiral phase transition.  Sec.~\ref{secspatialqq} covers
the temperature dependence of properties of spatial $\bar{q}q$ modes,
including masses and decays.  Finally we conclude in
Sec.~\ref{secsum}.  
}

\section{Dyson--Schwinger equations}
\label{secdse}
{\sloppy 
The dressed quark propagator $S(q)$ is the solution of the
DSE\footnote{We use Euclidean metric
$\{\gamma_\mu,\gamma_\nu\} = 2\delta_{\mu\nu}$,
$\gamma_\mu^\dagger = \gamma_\mu$ and
$a\cdot b = \sum_{i=1}^4 a_i b_i$.}
\begin{eqnarray}
\label{gendse}
\lefteqn{ S(p)^{-1} = Z_2\,i\,/\!\!\!p + Z_4\,m(\mu) }
\nonumber\\
        &+& Z_1 \int^\Lambda_q \!\! g^2 D_{\mu\nu}(k)
     \textstyle{\frac{\lambda^i}{2}}\gamma_\mu S(q)\Gamma^i_\nu(q,p) \,,
\end{eqnarray}
where $D_{\mu\nu}(k=p-q)$ is the renormalized dressed-gluon propagator,
and $\Gamma^i_\nu(q,p)$ is the renormalized dressed quark-gluon vertex.
The notation \mbox{$\int^\Lambda_q \equiv \int^\Lambda d^4 q/(2\pi)^4$}
stands for a translationally invariant regularization of the integral,
with $\Lambda$ being the regularization mass-scale.  The regularization
can be removed at the end of all calculations, and after
renormalization, by taking the limit \mbox{$\Lambda \to \infty$}.  The
solution of Eq.~(\ref{gendse}) is renormalized according to $S(p)^{-1} =
i\,/\!\!\!p + m(\mu)$ at a sufficiently large spacelike $\mu^2$, with
$m(\mu)$ the renormalized quark mass at the scale $\mu$.  In
Eq.~(\ref{gendse}), $S$, $\Gamma^i_\mu$ and $m(\mu)$ depend on the quark
flavor, although we have not indicated this explicitly.  The
renormalization constants $Z_2$ and $Z_4$ depend on the renormalization
point and the regularization mass-scale, but not on flavor: in our
analysis we employ a flavor-independent renormalization scheme.

The Bethe--Salpeter amplitude [BSA] for a meson bound state of a quark
$a$ having momentum $q_+$ and an antiquark $b$ having momentum $q_-$
is denoted by $\Gamma^{a\bar{b}}(q_+,q_-)$.  The meson momentum is
\mbox{$P= q_+ - q_-$} and satisfies $P^2 = -m^2$.  The BSAs are
solutions of the homogeneous BSE
\begin{equation}
\Gamma^{a\bar{b}}(p_+,p_-) =
        \int^\Lambda_{q} \!
        K(p,q;P) \,\chi^{a\bar{b}}(q_+,q_-) \,,
\label{homBSE}
\end{equation}
which has solutions at discrete values of $P^2$ only.  Here, $K$ is
the renormalized $\bar q q$ scattering kernel that is irreducible with
respect to a pair of $\bar q q$ lines, and
\mbox{$\chi^{a\bar{b}}(q,q') = S^a(q) \Gamma^{a\bar{b}}(q,q')
S^b(q')$} is the Bethe--Salpeter wavefunction.  In order to implement
Eq.~(\ref{homBSE}) while preserving the constraint $P= $
\mbox{$q_+ - q_-$}, it is necessary to specify how the total momentum 
is partitioned between the quark and the antiquark.  The general
choice is \mbox{$q_+ = q+\eta P$} and \mbox{$q_- = q - (1-\eta) P$}
where $\eta$ is the momentum partitioning parameter.  The choice of
$\eta$ corresponds to the choice of relative momentum; physical
observables should not depend on the choice.  This provides us with a
convenient check on the accuracy of the numerics in an actual model
calculation.

The meson BSA $\Gamma^{a\bar{b}}$ is normalized according to
the canonical normalization condition
\begin{eqnarray}
 \lefteqn{2\, P_\mu = N_c \; \frac{\partial}{\partial P_\mu} \Bigg\{ }
\nonumber\\ &&
       \int^\Lambda_q \! {\rm tr}_{\rm s}\, \big[
        \bar\Gamma^{a\bar{b}}(\tilde{q}',\tilde{q})\,S^a(q_+)\,
        \Gamma^{a\bar{b}}(\tilde{q},\tilde{q}')\,S^b(q_-) \big]
\nonumber\\ & + &
        \int^\Lambda_{k,q} \! {\rm tr}_{\rm s}\, \big[
        \bar\chi^{a\bar{b}}(\tilde{k}',\tilde{k})\,K(k,q;P)\,
        \chi^{a\bar{b}}(\tilde{q},\tilde{q}') \big] \Bigg\} \, ,
\nonumber\\ {}
\label{gennorm}
\end{eqnarray}
at the meson mass shell ${P^2=Q^2=-m^2}$, with 
\mbox{$\tilde{q}=q+\eta Q$},
\mbox{$\tilde{q}'=q-(1-\eta) Q$}, and similarly for $\tilde{k}$ and
$\tilde{k}'$.  The notation ${\rm tr}_{\rm s}$ denotes a trace over
Dirac spin.  For the antimeson BSA we have used the notation
\mbox{$\bar{\Gamma}(q_-,q_+) =$} \mbox{$ [C^{-1} \Gamma(-q_+,-q_-)
C]^{\rm t}$} in which $C=\gamma_2 \gamma_4$ is the charge conjugation
matrix, and $X^{\rm t}$ denotes the matrix transpose of $X$.

For pseudoscalar bound states the BSA is commonly decomposed
into~\cite{MR97}
\begin{eqnarray}
\label{genpion}
\lefteqn{ \Gamma(k+\eta P,k-(1-\eta)P) = }
\nonumber \\
        && \gamma_5 \big[ i E + \;/\!\!\!\! P \, F + 
        \,/\!\!\!k \, G + \sigma_{\mu\nu}\,k_\mu P_\nu \,H \big]\,,
\end{eqnarray}
where the invariant amplitudes $E$, $F$, $G$ and $H$ are Lorentz scalar
functions $f(k^2,k\cdot P;\eta)$.  The dependence of the amplitudes
upon \mbox{$k\cdot P$} can be conveniently represented by the following
expansion based on Chebyshev polynomials
\begin{equation}
\label{expansion}
\label{chebmom}
 f(k^2, k\cdot P) = \sum_{i=0}^\infty
        U_i(\cos\theta)\, (k P)^i\, f_i(k^2) \,,
\end{equation}
where \mbox{$\cos\theta = k\cdot P /(k P)$}.  For charge-parity
eigenstates such as the pion, each amplitude $E$, $F$, $G$, and $H$
will have a well-defined parity in the variable $k\cdot P$ if one
chooses $\eta=\frac{1}{2}$.  In this case, these amplitudes are either
entirely even ($E$, $F$, and $H$) or odd ($G$) in $k\cdot P$, and only
the even ($E$, $F$, and $H$) or odd ($G$) Chebyshev moments $f_i$ are
needed for a complete description.
}

\section{Ladder-rainbow model}
\label{secladderrainbow}
{\sloppy
We employ a model that has been developed recently for an efficient
description of the masses and decay constants of the light
pseudoscalar and vector mesons~\cite{MR97,MT99}.  This consists of the
rainbow truncation of the DSE for the quark propagator and the ladder
truncation of the BSE for the meson BSAs.  The required effective
$\bar q q$ interaction is constrained by perturbative QCD in the
ultraviolet and has a phenomenological infrared behavior.  In
particular, the rainbow truncation of the quark DSE,
Eq.~(\ref{gendse}), is
\begin{equation}
\label{ourDSEansatz}
Z_1 \, g^2 D_{\mu \nu}(k)\, \Gamma^i_\nu(q,p) \rightarrow
 {\cal G}(k^2)\, D_{\mu\nu}^{\rm free}(k)\, \gamma_\nu
                                        \textstyle\frac{\lambda^i}{2} \,,
\end{equation}
where $D_{\mu\nu}^{\rm free}(k=p-q)$ is the free gluon propagator in
Landau gauge.  The consistent ladder truncation of the BSE,
Eq.~(\ref{homBSE}), is
\begin{equation}
\label{ourBSEansatz}
        K(p,q;P) \to
        -{\cal G}(k^2)\, D_{\mu\nu}^{\rm free}(k)
        \textstyle{\frac{\lambda^i}{2}}\gamma_\mu \otimes
        \textstyle{\frac{\lambda^i}{2}}\gamma_\nu \,,
\end{equation}
where \mbox{$k=p-q$}.  These two truncations are consistent in the
sense that the combination produces vector and axial-vector vertices
satisfying the respective WTI.  In the axial case, this ensures that
in the chiral limit the ground state pseudoscalar mesons are the
massless Goldstone bosons associated with chiral symmetry breaking;
with nonzero current quark masses it leads to a generalisation of the
Gell-Man--Oaks--Renner relation~\cite{MRT98}.  In the vector
case, this ensures conservation of the meson electromagnetic current
in impulse approximation as described in the next section.

The model is completely specified once a form is chosen for the
``effective coupling'' ${\cal G}(k^2)$.  The ultraviolet behavior is
chosen to be that of the QCD running coupling $\alpha(k^2)$; the
ladder-rainbow truncation then generates the correct perturbative QCD
structure of the DSE-BSE system of equations.  The phenomenological
infrared form of ${\cal G}(k^2)$ is chosen so that the DSE kernel
contains sufficient infrared enhancement to produce an empirically
acceptable amount of dynamical chiral symmetry breaking as represented
by the chiral condensate~\cite{HMR98}.

We employ the Ansatz found to be successful for the light mesons in
Refs.~\cite{MR97,MT99}
\begin{eqnarray}
\label{gvk2}
\frac{{\cal G}(k^2)}{k^2} &=&
        \frac{4\pi^2\, D \,k^2}{\omega^6} \, {\rm e}^{-k^2/\omega^2}
\nonumber \\  &+&
 \frac{ 4\pi^2\, \gamma_m \; {\cal F}(k^2)}
        {\case{1}{2} \ln\left[\tau +
        \left(1 + k^2/\Lambda_{\rm QCD}^2\right)^2\right]} \,,
\end{eqnarray}
with \mbox{$\gamma_m=\frac{12}{33-2N_f}$} and 
\mbox{${\cal F}(s)=(1 - \exp(\frac{-s}{4 m_t^2}))/s$}.  The first 
term implements the strong infrared enhancement in the region 
\mbox{$0 < k^2 < 1\,{\rm GeV}^2$} required for sufficient dynamical
chiral symmetry breaking.  The second term serves to preserve the
one-loop renormalization group behavior of QCD.  We use
\mbox{$m_t=0.5\,{\rm GeV}$}, \mbox{$\tau={\rm e}^2-1$},
\mbox{$N_f=4$}, and we take \mbox{$\Lambda_{\rm QCD} = 0.234\,{\rm
GeV}$}.  The renormalization scale is chosen to be 
\mbox{$\mu=19\,{\rm GeV}$} which is well into the domain where
one-loop perturbative behavior is appropriate~\cite{MR97,MT99}.  The
remaining parameters, \mbox{$\omega =$} \mbox{$0.4\,{\rm GeV}$} and
\mbox{$D=0.93$} \mbox{${\rm GeV}^2$} along with the quark masses, 
are fitted to give a good description of $\langle\bar q q\rangle$,
$m_{\pi/K}$ and $f_{\pi}$.  The subsequent values for $f_K$ and the
masses and decay constants of the vector mesons $\rho, \phi, K^\star $
are found to be within 10\% of the experimental data~\cite{MT99}, see
Table~\ref{sumres}.  A detailed analysis of the relationship between
QCD and this Landau gauge, rainbow-ladder truncation of the DSEs with
renormalization group improvement, can be found in the originating
work~\cite{MR97}.  A comparison with lattice QCD
results~\cite{tonylatticequark} confirms the qualitative behavior for
the dressed quark propagator produced by this
model~\cite{Maris:ViennaConf,Tandy:Adel2001}.
\TABLE[ht]{
\caption{\label{sumres}
Overview of the results of the model for the meson masses and decay
constant, adapted from Ref.~\protect\cite{MT99}.  The experimental
value for the condensate is taken from Ref.~\protect\cite{Lein97},
the other experimental data are from Ref.~\protect\cite{PDG}. }
\begin{tabular}{l|ccr}
        & \multicolumn{1}{r}{experiment}
        & \multicolumn{1}{r}{calculated}        &\\
        & \multicolumn{1}{r}{(estimates)}
        & \multicolumn{1}{r}{($^\dagger$ fitted)} &\\ \hline
$m^{u=d}_{\mu=1 {\rm GeV}}$ &
        \multicolumn{1}{r}{ 5 - 10 MeV}  &
        \multicolumn{1}{r}{ 5.5 MeV}     &\\
$m^{s}_{\mu=1 {\rm GeV}}$ &
        \multicolumn{1}{r}{ 100 - 300 MeV} &
        \multicolumn{1}{r}{ 125 MeV   }    &\\ \hline
- $\langle \bar q q \rangle^0_{\mu}$
                & (0.236 GeV)$^3$ & (0.241$^\dagger$)$^3$ &\\
$m_\pi$         &  0.1385 GeV &   0.138$^\dagger$ &\\
$f_\pi$         &  0.131 GeV &   0.131$^\dagger$ &\\
$m_K$           &  0.496 GeV  &   0.497$^\dagger$ &\\
$f_K$           &  0.160 GeV  &   0.155        &\\ \hline
$m_\rho$        &  0.770 GeV  &   0.742        &\\
$f_\rho$        &  0.216 GeV  &   0.207        &\\
$m_{K^\star}$   &  0.892 GeV  &   0.936        &\\
$f_{K^\star}$   &  0.225 GeV  &   0.241        &\\
$m_\phi$        &  1.020 GeV  &   1.072        &\\
$f_\phi$        &  0.236 GeV  &   0.259        &\\
\end{tabular}
}

Recent reviews~\cite{bastirev,Alkofer:2000wg,cdrESI} put this model in
a wider perspective.  These reviews include a compilation of results
for both meson and baryon physics with similar models, an analysis of
how quark confinement is manifest in solutions of the DSEs, and both
finite temperature and finite density extensions.  The question of the
accuracy of the ladder-rainbow truncation has also received some
attention; it was found to be particularly suitable for the flavor
octet pseudoscalar mesons and for the vector mesons, since the
next-order contributions in a quark-gluon skeleton graph expansion
have a significant amount of cancellation between repulsive and
attractive corrections~\cite{BRS96}.  }

\section{Electromagnetic form factors in impulse approximation}
\label{secff}
{\sloppy
The 3-point function describing the coupling of a photon with momentum
$Q$ to a pseudoscalar meson, with initial and final momenta 
\mbox{$P \pm Q/2$}, can be written as the sum of two terms
\begin{eqnarray}
\label{mesonff}
\Lambda^{a\bar{b}}_\nu(P,Q) &=&
                \hat{Q}^a \, \Lambda^{a\bar{b}a}_\nu
                + \hat{Q}^{\bar{b}} \, \Lambda^{a\bar{b}\bar{b}}_\nu \,,
\end{eqnarray}
where $\hat{Q}$ is the quark or antiquark electric charge, and where
$\Lambda^{a\bar{b}a}(P,Q)$ and $\Lambda^{a\bar{b}\bar{b}}(P,Q)$
describe the coupling of a photon to the quark ($a$) and antiquark
($\bar{b}$) respectively.  The meson form factor is defined by
\begin{equation}
 \Lambda^{a\bar{b}}_\nu(P,Q) =
        2\;P_\nu\;F(Q^2) \,,
\end{equation}
with the charge radius defined by \mbox{$r^2 = -6 F'(0)$}.
Analogously, we can define a form factor for each of the two terms on
the RHS of Eq.~(\ref{mesonff}), for example
\begin{equation}
 \Lambda^{a\bar{b}\bar{b}}_\nu(P,Q) =
        2\;P_\nu\;F_{a\bar{b}\bar{b}}(Q^2) \,.
\end{equation}
Current conservation dictates that each of the form factors
$F_{a\bar{b}\bar{b}}(Q^2)$ and $F_{a\bar{b}a}(Q^2)$ is one at
\mbox{$Q^2=0$}.

The impulse approximation allows form factors to be described in terms
of dressed quark propagators, bound state BSAs, and the dressed
$qq\gamma$-vertex.  We denote by $\Gamma^a_\mu(q_+,q_-)$ the vertex
describing the coupling of a photon with momentum \mbox{$Q=q_+ - q_-$}
to a quark of flavor $a$ with final and initial momenta $q_+$ and
$q_-$ respectively.  With this notation, the vertices in
Eq.~(\ref{mesonff}) take the form~\cite{MTpiK00}
\begin{eqnarray}
\lefteqn{\Lambda^{a\bar{b}\bar{b}}_\nu(P,Q) =
        N_c \int^\Lambda_k \!{\rm tr}_{\rm s}\, \big[
        S^a(q) \, \Gamma^{a\bar{b}}(q,q_+) \, S^b(q_+) }
\nonumber \\ && {\;\;\;\;} \times
        i \Gamma^{b}_\nu(q_+,q_-)\, S^b(q_-) \,
        \bar\Gamma^{a\bar{b}}(q_-,q) \big] \;,
        \;\;\;\;\;\;\;\;\;\;\;\;\;\;\;\;
\label{triangle}
\end{eqnarray}
where \mbox{$q = k+P/2$} and \mbox{$q_\pm = k-P/2 \pm Q/2$}.
The expression for $\Lambda^{a\bar{b}a}_\nu$ is analogous.
}

\subsection{The quark-photon vertex}
{\sloppy
The quark-photon vertex is the solution of the renormalized
inhomogeneous BSE with the same kernel $K$ as the homogeneous BSE for
meson bound states.  That is, for photon momentum $Q$,
\begin{eqnarray}
\Gamma^a_\mu(p_+,p_-) &=& Z_2\, \gamma_\mu +
        \int^\Lambda_q  K(p,q;Q)\, S^a(q_+)
\nonumber \\ && {}\times
        \Gamma^a_\mu(q_+,q_-) \, S^a(q_-) \,,
\label{verBSE}
\end{eqnarray}
with $p_\pm = p \pm Q/2$ and $q_\pm= q\pm Q/2$.  Because of gauge
invariance, it satisfies the WTI
\begin{equation}
i\,Q_\mu \,\Gamma^a_{\mu}(p_+,p_-)  =
                       S_a^{-1}(p_+) - S_a^{-1}(p_-) \,.
\label{wtid}
\end{equation}

The general form of the quark-photon vertex $\Gamma^a_\mu(q;Q)$ requires
a decomposition into twelve independent Lorentz covariants, which
can be made from
the three vectors $\gamma_\mu$, the relative momentum $q_\mu$, and the
photon momentum $Q_\mu$, each multiplied by one of the four
independent matrices $1\hspace{-3pt}$l, $\,/\!\!\!q$, $\;/\!\!\!\!Q$,
and $\sigma_{\mu\nu} q_\mu Q_\nu$.  Four of the covariants represent
the longitudinal components which are completely specified by the WTI
in terms of the quark propagator and they do not contribute to elastic
form factors.  The solution of the BSE for the transverse vertex can
be expanded in eight covariants $T^i_\mu(q;Q)$ with the corresponding
amplitudes being Lorentz scalar functions.  The choice of covariant
basis $T^i_\mu(q;Q)$ is constrained by the required properties under
Lorentz and CPT transformations, but is not unique.  A convenient
basis that facilitates projection of the BSE, Eq.~(\ref{verBSE}), has
been given in Ref.~\cite{MT99pion} and a discussion of the results for
the vertex can also be found there.

Note that solutions of the homogeneous version of Eq.~(\ref{verBSE})
define vector meson bound states with masses \mbox{$m_V^2=-Q^2$} at
discrete timelike momenta $Q^2$.  It follows that $\Gamma^a_\mu$ has
poles at those locations and, in the vicinity of the bound states,
behaves like~\cite{MT99pion}
\begin{equation}
 \Gamma^a_\mu(p_+,p_-) \rightarrow
 \frac{\Gamma_\mu^{a\bar{a}\,V}(p_+,p_-) f_V m_V}{Q^2 + m_V^2} \; ,
\label{comres}
\end{equation}
where $\Gamma_\mu^{a\bar{a}\,V}$ is the $a\bar{a}$ vector meson BSA, and
$f_V$ is the electroweak decay constant.  For the photon coupled to $u$-
and $d$-quarks, this results in a $\rho$-meson pole at $Q^2 = -0.6\,{\rm
GeV}^2$.  We will come back to this behavior in the timelike region in
Sec.~\ref{sectimelike}.
}

\subsection{Current conservation}
{\sloppy 
At \mbox{$Q=0$} the quark-photon vertex is completely specified by the
differential Ward identity
\begin{equation}
i \,\Gamma^b_{\mu}(p,p) =
        \frac{\partial}{\partial p_\mu} S_b^{-1}(p) \; .
\label{wid}
\end{equation}
If this is inserted in Eq.~(\ref{triangle}), one finds after a change
of integration variables $k \rightarrow q-P/2$
\begin{eqnarray}
\lefteqn{ \Lambda^{a\bar{b}\bar{b}}_\nu(P,0) =
        2 \, P_\mu \, F_{a\bar{b}\bar{b}}(0) = }
\nonumber \\  &&
        N_c \, \int^\Lambda_q \! {\rm tr}_{\rm s}\, \bigg[
        \bar\Gamma^{a\bar{b}}(q',q) \, S^a(q) \,
        \Gamma^{a\bar{b}}(q,q') \,
        \frac{\partial S^b(q')}{\partial P_\mu} \bigg]\,,
\nonumber \\ {}
\end{eqnarray}
with $q'=q-P$.  Comparing this expression to Eq.~(\ref{gennorm}) with
$\eta = 0$, we recognize that the physical result \mbox{$F(Q^2=0)=1$}
follows directly from the canonical normalization condition for
$\Gamma^{a\bar{b}}$ if the BSE kernel $K$ is independent of the meson
momentum $P$ and the quark-photon vertex satisfies the differential
Ward identity~\cite{roberts96}.  

With an arbitrary value for the momentum partitioning parameter
$\eta$, the relation between the normalization condition for
$\Gamma^{a\bar{b}}$ and electromagnetic current conservation is not so
straightforward.  A shift in the value of $\eta$ in loop diagrams
(without external quark lines) is equivalent to a shift in integration
variables.  For processes that are not anomalous, loop integrals are
independent of a shift in integration variables provided that all
approximated quantities in the integrand respect Poincar\'e
covariance.  In this respect, particular attention must be paid to the
representation and approximation used for the BSAs.  The general BSE
vertex solution $\Gamma(q_+,q_-)$, considered as a function of the
incoming and outgoing quark momenta, does not depend on $\eta$.  It is
only in commonly used decompositions such as Eq.~(\ref{genpion}), that
the value of $\eta$ becomes relevant.  The amplitudes $E$, $F$, $G$,
and $H$ are scalar functions of $k^2$ and $k\cdot P$ which {\em do}
depend on $\eta$, i.e., on the choice of a relative momentum.  Under a
change of $\eta$, some of the different Dirac covariants and
associated amplitudes (such as $F$ and $G$) will mix, as will the
various Chebyshev moments $f_i$ in Eq.~(\ref{chebmom}).  However, the
net result from use of a complete representation of both the Dirac
matrix structure and the $k\cdot P$ variable is independent of $\eta$
and choice of integration variables in loop integrals; this has been
checked explicitly for the decay constants~\cite{MR97} and for form
factors~\cite{MTpiK00}.

The ladder BSE kernel is independent of the meson momentum $P$, and
with quark propagators dressed in rainbow approximation and the
dressed quark-photon vertex produced in ladder approximation, the WTI
is satisfied~\cite{MT99pion}; thus impulse approximation for the
charge form factor in combination with rainbow-ladder truncation is
ideal in the sense that the resulting meson electromagnetic current is
conserved.  That is, the correct electric charge of the
meson is produced independent of the model parameters, and independent
of the momentum partitioning parameter $\eta$.  
}

\subsection{Beyond rainbow-ladder truncation}
{\sloppy
If one goes beyond the rainbow-ladder truncation for the DSEs for the
propagators, BSAs and quark-photon vertex, one has to go beyond
impulse approximation for the form factors to ensure current
conservation~\cite{MTpiK00}.  For example, one could include
corrections to the rainbow-ladder DSE and BSE kernels that are
higher-order in $\alpha_s$, as depicted in Fig.~\ref{DSEbeyond}.
Following the general procedure developed in Ref.~\cite{BRS96}, one
can show that both the WTI, Eq.~(\ref{wtid}), and the differential
Ward identity, Eq.~(\ref{wid}), are preserved in the truncation
indicated in Fig.~\ref{DSEbeyond}.  Also preserved is the axial-vector
WTI, which is important for the Goldstone nature of the
pions~\cite{MRT98}.
%
\FIGURE[ht]{
\epsfig{figure=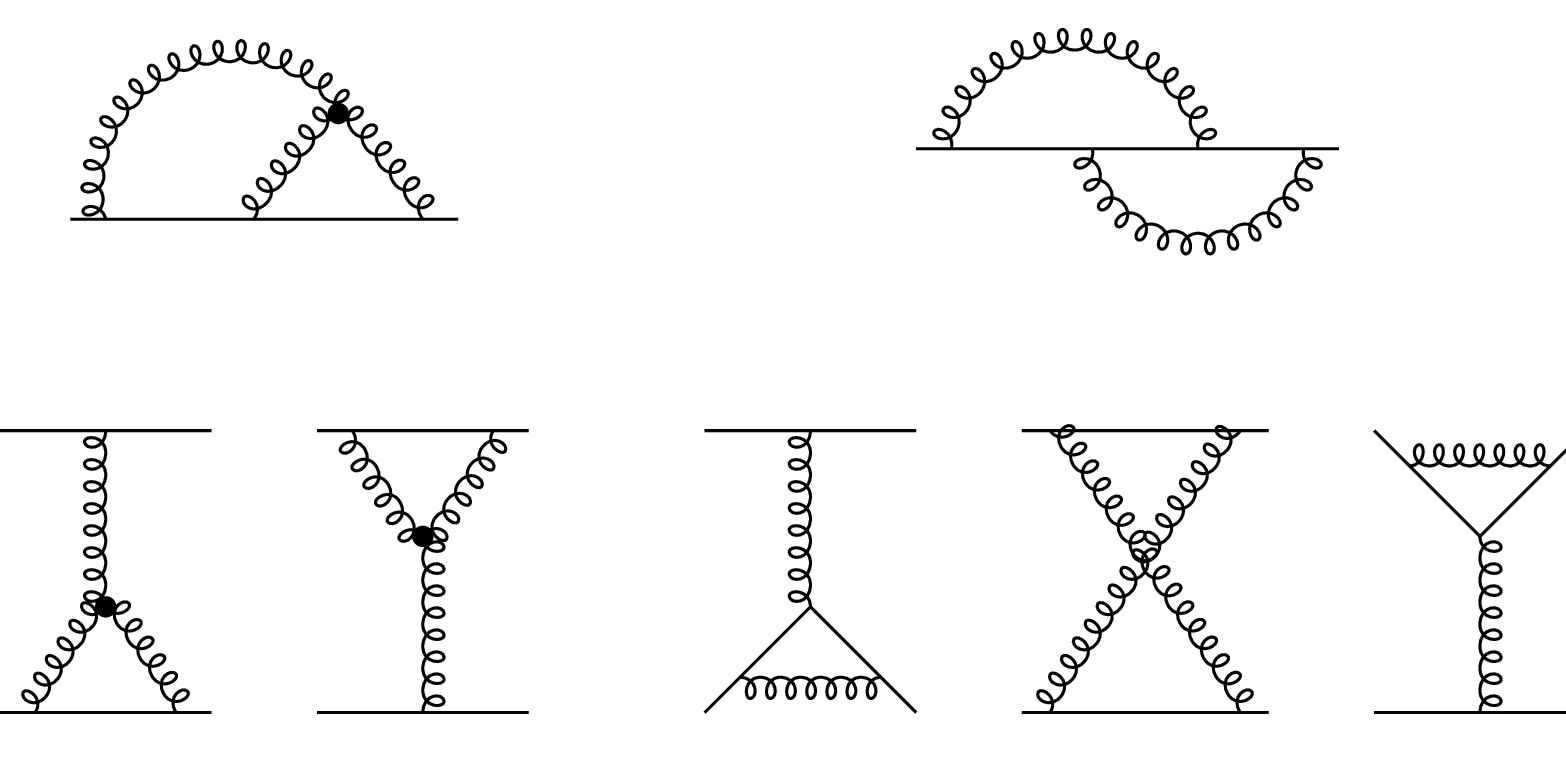, height=2.5cm}
\caption{
The two leading-order vertex corrections to the rainbow DSE (top) and
the corresponding five diagrams to be added to the ladder BSE kernel
(bottom) for consistency with the relevant WTIs.  The quark and gluon
lines indicate dressed propagators in this and the subsequent figures.
\label{DSEbeyond} }
}
%

The resulting BSE kernel $K(q,p;P)$ now becomes dependent on the meson
momentum $P$, which means that the second term of the normalization
condition, Eq.(\ref{gennorm}), is nonzero.  To be specific, with the
choice $\eta = 0$, this introduces the four extra terms in the
normalization condition depicted in Fig.~\ref{normbeyond}.  These four
additional diagrams are generated from the BSE kernel in the bottom
part of Fig.~\ref{DSEbeyond} by taking the derivative with respect to
the meson momentum $P$, where $P$ flows through one quark propagator
only.
%
\FIGURE[ht]{
\epsfig{figure=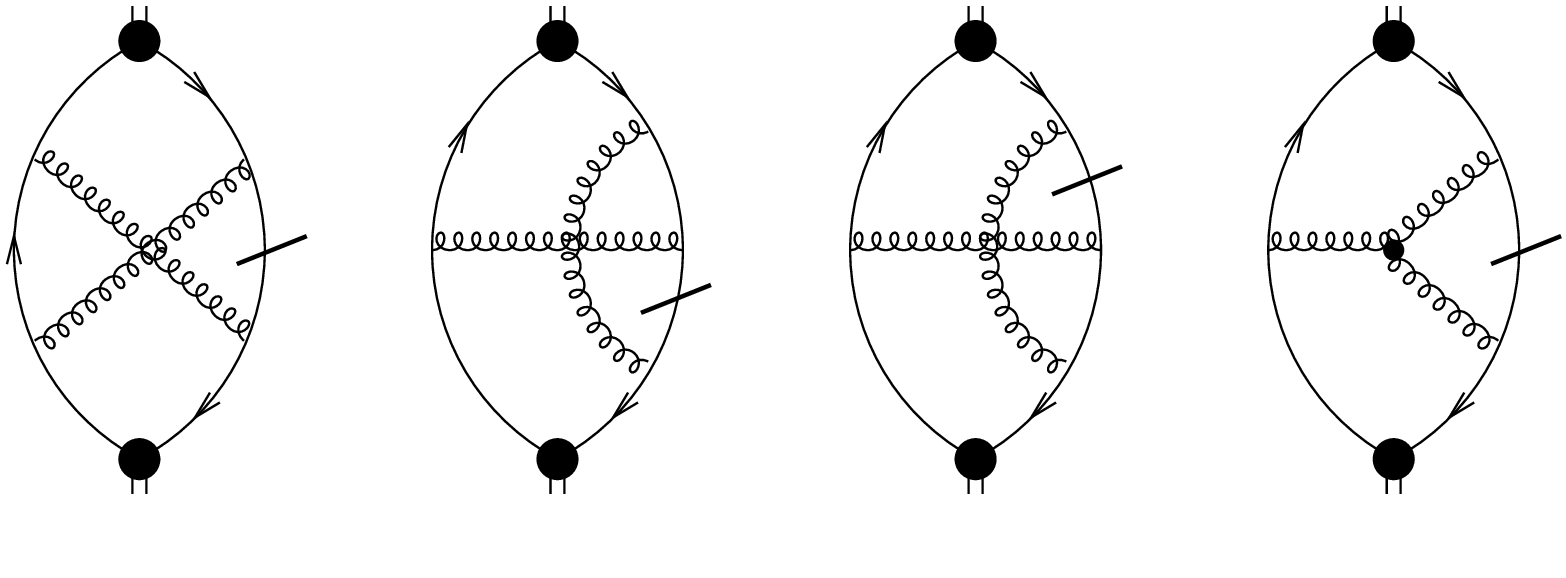, height=2.5cm}
\caption{
The four diagrams arising from the $P$-dependence of the kernel in the
normalization condition, Eq.~(\protect\ref{gennorm}), if one includes
the diagrams of Fig.~\protect\ref{DSEbeyond} into the DSE dynamics,
and chooses all of the meson momentum $P$ to flow through one quark
only.  The derivatives with respect to $P$ are marked by slashes.
\label{normbeyond} }
}
%
Since a derivative with respect to $P$ is equivalent to the insertion
of a zero-momentum photon according to the differential WTI,
Eq.~(\ref{wtid}), it is obvious which diagrams have to be added to the
impulse approximation for the vertex $\Lambda^{a\bar{b}\bar{b}}$ to
maintain current conservation~\cite{MTpiK00}, and they are displayed
in Fig.~\ref{tribeyond}.  In the limit $Q\rightarrow 0$ these four
additional diagrams become identical to the four additional diagrams
in Fig.~\ref{normbeyond}, provided that the vertex satisfies the
differential WTI.  Of course, there are similar contributions to
$\Lambda^{a\bar{b}a}$, which can be identified with terms in the
normalization condition with $\eta = 1$.
%
\FIGURE[ht]{
\epsfig{figure=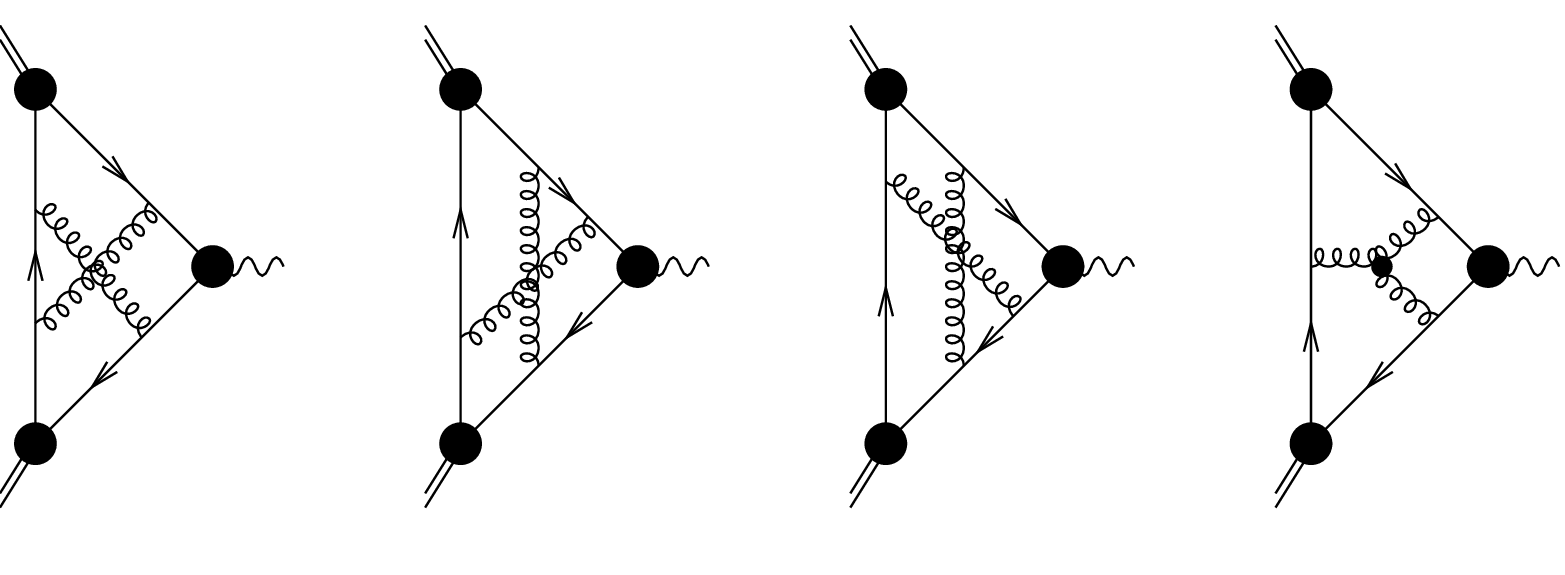, height=2.5cm}
\caption{
The four corrections to the impulse approximation,
Eq.~(\protect\ref{triangle}), necessary to maintain current
conservation if one includes the diagrams of
Fig.~\protect\ref{DSEbeyond} into the DSE dynamics. 
\label{tribeyond} }
}
%

A related topic is the correction to the impulse approximation from
pion and kaon loops.  Simple addition of such loops without
supplementing the ladder-rainbow truncation for the DSEs, will
generally violate current conservation.  A consistent treatment of the
kernels for both the DSE and BSE equations and the approximation for
the photon-hadron coupling is necessary for current conservation.  At
present it is not clear how to incorporate meson loops
self-consistently in such an approach, but we expect corrections
coming from such loops to be small in the spacelike region.  In
Ref.~\cite{ABR95} it was demonstrated that the dressed quark core can
generate most of the pion charge radius, and that pion loops
contribute less than 15\% to $r_\pi^2$.  For larger values of $Q^2$
the effect from meson loops reduces even further, and for 
\mbox{$Q^2 > 1 \,{\rm GeV}^2$} we expect the contribution of such 
loops to be negligible.
}

\subsection{Results for the $\pi$ and $K^{0,\pm}$ form factors}
\label{secnumcalcff}
{\sloppy
The pion and kaon form factors are given by
\begin{eqnarray}
\label{fpi}
 F_{\pi}(Q^2) = \case{2}{3}F_{u\bar{d}u}(Q^2)
        + \case{1}{3}F_{u\bar{d}\bar{d}}(Q^2)            \,,\\
\label{fKplus}
 F_{K^+}(Q^2) = \case{2}{3}F_{u\bar{s}u}(Q^2)
        + \case{1}{3}F_{u\bar s \bar s}(Q^2)            \,,\\
\label{fK0}
 F_{K^0}(Q^2) = -\case{1}{3}F_{d\bar{s}d}(Q^2)
        + \case{1}{3}F_{d\bar s \bar s}(Q^2)            \,,
\end{eqnarray}
where the quark and antiquark charges are evident.  It is only
recently that a consistent implementation of the impulse approximation
for these form factors has been carried out with solutions of the
ladder-rainbow DSEs for each required element: propagators, meson
BSAs, and the photon-quark vertex~\cite{MTpiK00,MT99pion}.
Non-analytic effects from vector mesons are automatically taken into
account, because these vector $\bar{q} q$ bound states appear as poles
in the quark-photon vertex solution~\cite{MT99pion}.

We employ the model described in Sec.~\ref{secladderrainbow}, without
any parameter adjustment.  Since we work in the SU(2) isospin limit,
$u$ and $d$ quarks are identical apart from their charge,
and we have for the pion electromagnetic form factor simply
\mbox{$F_{\pi}(Q^2) = F_{u\bar{u}u}(Q^2)$}.  Thus there are only three 
independent form factors, $F_{u\bar{u}u}(Q^2)$, $F_{u\bar{s}u}(Q^2)$,
and $F_{u\bar s\bar s}(Q^2)$.  With this model we have demonstrated
explicitly that the impulse approximation is indeed independent of the
unphysical parameter $\eta$ and satisfies current conservation,
provided that complete representations of the Dirac matrix structure
and the $k\cdot P$ dependence are used~\cite{MTpiK00}.  However, with
the leading terms in the Chebyshev expansion of the BSAs and the
quark-photon vertex given in Eq.~(\ref{expansion}) only, current
conservation is violated, and the results do depend on the momentum
routing in the loop integral.  Including terms up to order $(k\cdot
P)^3$ in this expansion restores current conservation and the
independence of the momentum routing to within the numerical accuracy
which we estimate to be around 1\%.  Higher order terms do not change
the results more than 1\%, at least not at the values of $Q^2$ we have
considered.  Also a truncation of the BSAs or the quark-photon vertex
in terms of their Dirac amplitudes ($E$, $F$, $G$, and $H$ for the
pseudoscalar mesons, and eight transverse amplitudes for the
quark-photon vertex) generally leads to a violation of current
conservation and/or independence of the choice of integration
variable.
%
\FIGURE[ht]{
\epsfig{figure=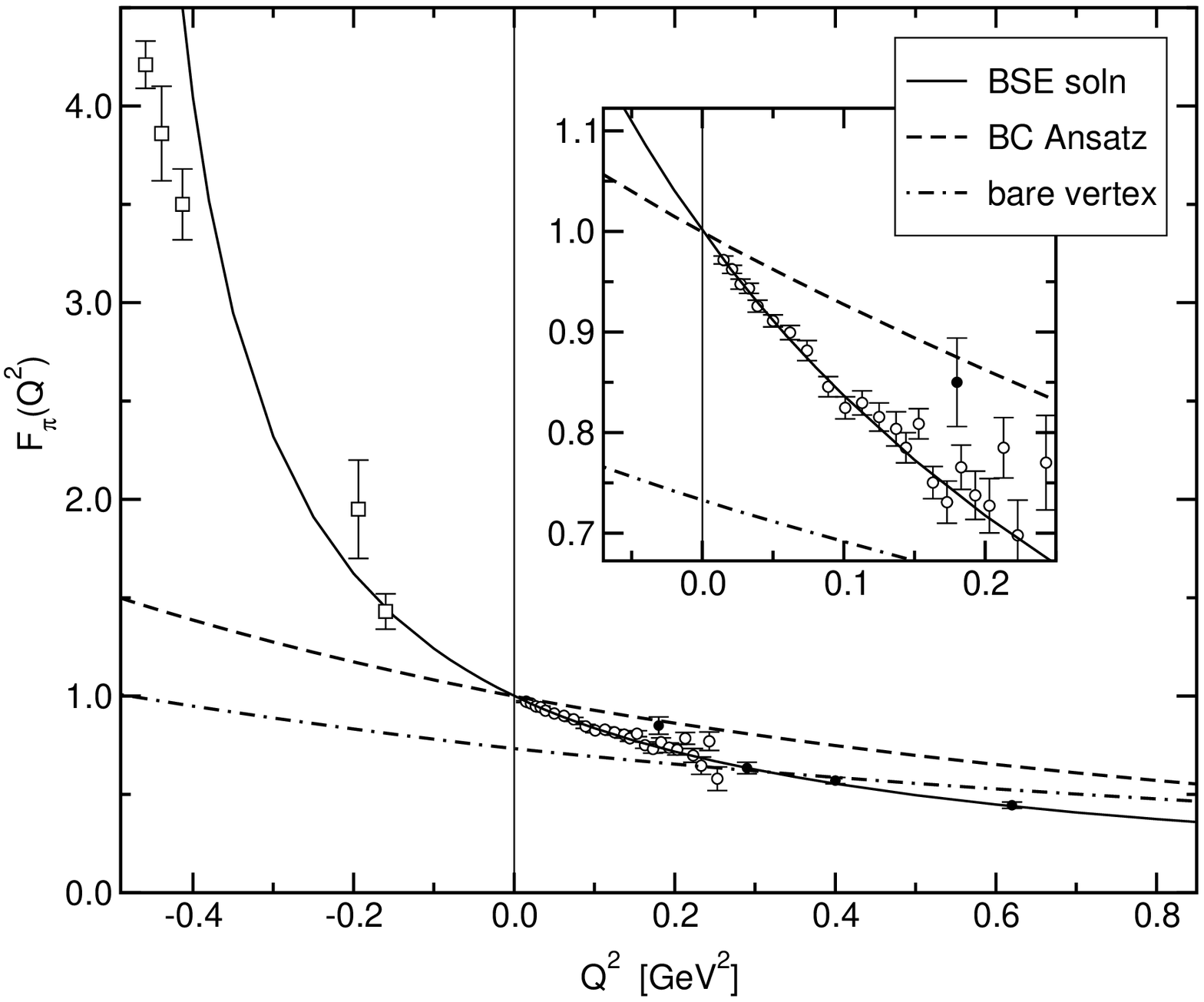,height=5.9cm} 
\caption{ 
The pion charge form factor $F_\pi(Q^2)$ as obtained from different
treatments of the quark-photon vertex.  The inset shows the $Q^2$
region relevant for the charge radius.  The data correspond to
$|F_\pi|$, taken from Refs.~\protect\cite{B76} (circles),
\protect\cite{B85} (squares), and \protect\cite{A86} (dots).
\label{fig:piFF} } }
%

In Fig.~\ref{fig:piFF} we show our results for the pion form
factor~\cite{MT99pion} around $Q^2 = 0$.  With dressed propagators,
the use of a bare quark-photon vertex in Eq.~(\ref{triangle}) clearly
violates electromagnetic current conservation and leads to $F_\pi(0)
\neq 1$. Use of the Ball--Chiu Ansatz~\cite{BC80} for the dressed
quark-photon vertex conserves the electromagnetic current and ensures
$F(Q^2=0)=1$.  However, the behavior of the form factor away from
$Q^2=0$ is not constrained by current conservation, and in the present
model, use of the Ball--Chiu Ansatz leads to a value for $r_\pi^2$
which is about 50\% too small~\cite{MT99pion}.  With the quark-photon
vertex produced by solution of the ladder BSE, and with quark
propagators from the rainbow DSE, all constraints from current
conservation are satisfied and the calculated value of $r^2_\pi$ is
within 5\% of the experimental value~\cite{MT99pion}.  Since the
impulse approximation omits meson loop mechanisms such as pion
rescattering, previously estimated to add up to 15\% to the charge
radius~\cite{ABR95}, the precision of the agreement shown in
Fig.~\ref{fig:piFF} should be considered preliminary until these
aspects are investigated within the present model.

%
\FIGURE[ht]{
\epsfig{figure=fpiJlab.eps,height=5.5cm}
\caption{
Our results for the pion charge form factor~\protect\cite{MTpiK00}
compared to new data from JLab~\protect\cite{Volmer01} including a
re-analysis of older data. \label{fig:Fpi_Jlab} } 
}
%
In Fig.~\ref{fig:Fpi_Jlab} we show our results for $Q^2 F_\pi(Q^2)$ on
a larger $Q^2$-domain~\cite{MTpiK00}.  We obtain an excellent
description of the spacelike $Q^2$ data, including the latest JLab
results, without any adjustment of the model parameters.  Our result
on this $Q^2$ domain can quite well be described by a simple monopole,
with a mass slightly lower than $m_\rho$.  

%
\FIGURE[ht]{
\epsfig{figure=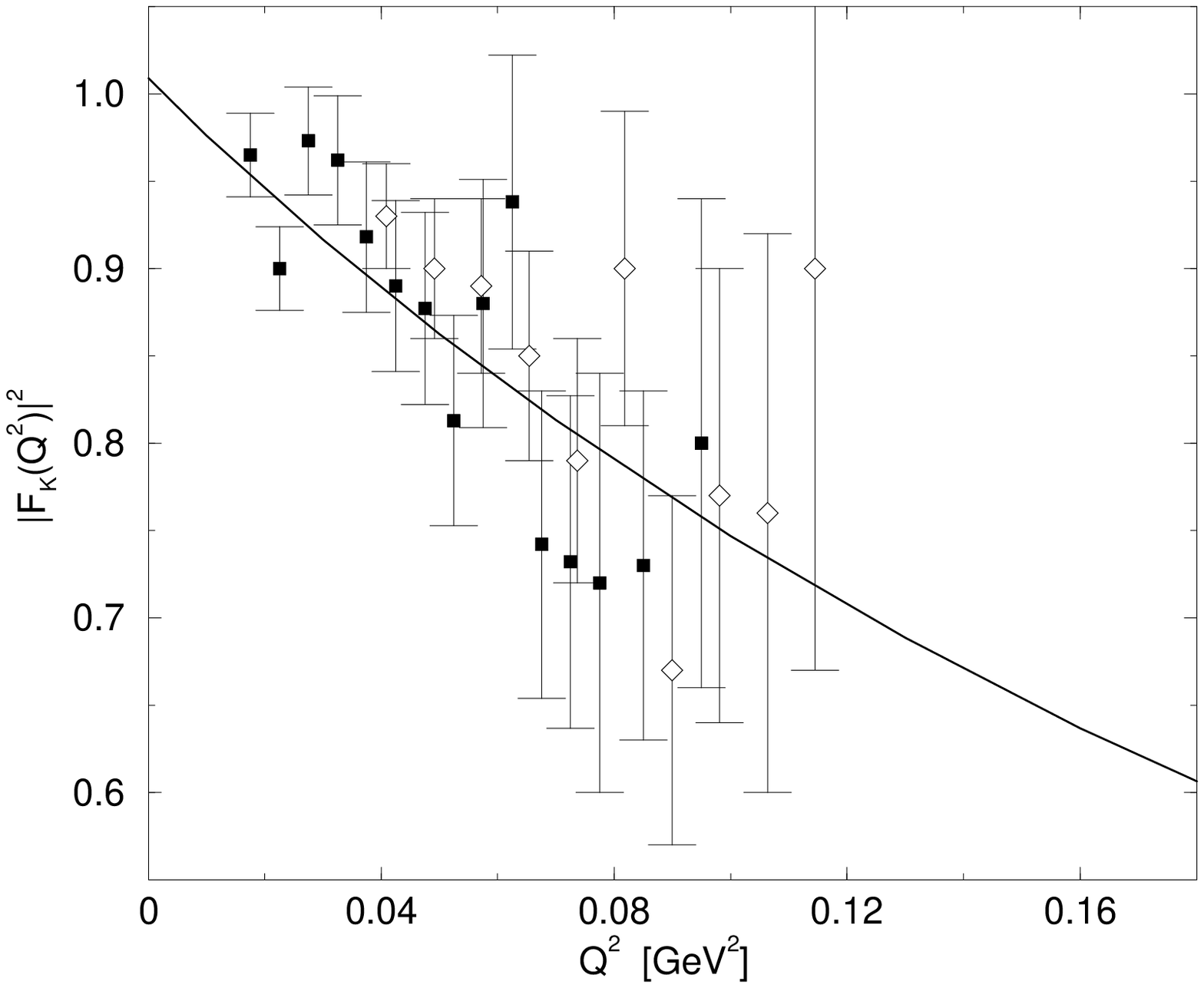, height=5.5cm}
\caption{
The calculated $K^+$ form factor compared to the data from
Refs.~\protect\cite{Dally80} (open diamonds) and \protect\cite{A86K}
(solid squares).  Within numerical errors, $F_{K^+}(0) = 1$.
\label{fig:Kplus} }
}
%
The obtained charged kaon form factor is also in good agreement with
the available data, see Fig.~\ref{fig:Kplus}.  The behavior of $F_K$
on a larger spacelike $Q^2$-domain anticipated for future
JLab~\cite{cebafK} data, is discussed elsewhere~\cite{MTpiK00}.  The
obtained charge radii for the pion and both the neutral and charged
kaons are presented in Table~\ref{resrK}, and agree well with the
experimental data.
\TABLE[ht]{
\caption{\label{resrK}
Our results for the charge radii, compared with the experimental values
given in Refs.~\protect\cite{A86,A86K,M78}.}
\begin{tabular}{l|ccr}
radii & experiment              & calculated &\\ \hline
$r_\pi^2$    & $ 0.44 \pm 0.01 $ fm$^2$&  0.45  fm$^2$ & \\
$r_{K^+}^2$  & $ 0.34 \pm 0.05 $ fm$^2$&  0.38  fm$^2$ & \\
$r_{K^0}^2$  & $-0.054\pm 0.026$ fm$^2$& $-$0.086 fm$^2$ &
\end{tabular}
}
These charge radii are somewhat larger than those
obtained in a previous study~\cite{BRT96} that was framed in terms of
semi-phenomenological representations for the confined quark
propagators and the BSAs within the impulse approximation.  The main
difference with that work is that here we use numerical solutions of
truncated DSEs for all the elements needed in Eq.~(\ref{triangle}),
and that all our parameters were fixed previously.
}

\subsection{Timelike behavior}
\label{sectimelike}
{\sloppy
Experimentally, $F_\pi(Q^2)$ shows a resonance peak at 
\mbox{$Q^2 = -m_\rho^2$} (in Euclidean metric, $Q^2<0$ corresponds to
the timelike region).  In general, for timelike photon momenta $Q^2$
in the vicinity of a vector meson mass-shell, the pseudoscalar meson
charge form factor $F_P(Q^2)$ will exhibit a resonance peak associated
with the propagation of intermediate state vector mesons followed by
the decay \mbox{$V \to P P$}.  That is~\cite{MT99pion}
\begin{equation}
 F_P(Q^2) \to \frac{ g_{VPP} \; m_V^2}
        { g_V \,  (Q^2 + m_V^2 - im_V \, \Gamma_V) } \;.
 \label{rhopole}
\end{equation}
Here \mbox{$m_V^2/g_V$} is the \mbox{$V$-$\gamma$} coupling strength
fixed by the \mbox{$V \to e^+ \, e^-$} decay, $g_{V PP}$ is the
coupling constant for the \mbox{$V \to PP$} decay, and $\Gamma_V$ is
the V meson width, which is dominated by the latter process.
%
\FIGURE[hb]{
\epsfig{figure=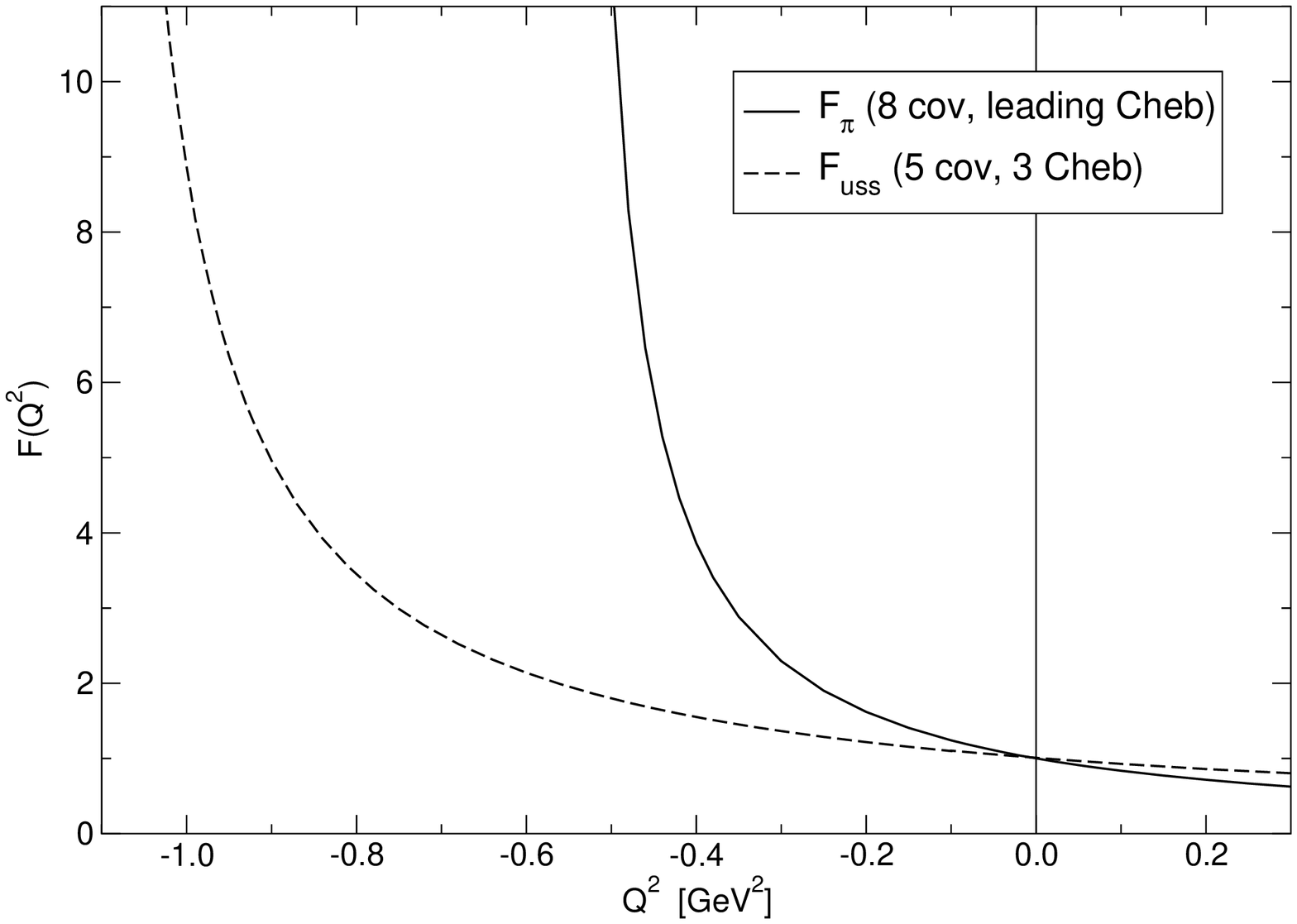, height=4.5cm}
\caption{Timelike behavior of the individual form factors that make up the
pion and kaon charge form factors. \label{timelikeFF} }
}
%

This vector meson mechanism in the charge form factors of the pion and
kaon is generated by the pole structure in the quark-photon vertex
exhibited in Eq.~(\ref{comres}); the relationship of the electroweak
decay constant $f_V$ in Eq.~(\ref{comres}) with $g_V$ is \mbox{$f_\rho
m_\rho = \sqrt{2} m_\rho^2 /g_\rho$} for the $\rho^0$, and
\mbox{$f_\phi m_\phi = 3 \, m_\phi^2 /g_\phi$} for the
$\phi$~\cite{MT99}.  In the present study, the pole is at a real
timelike value of $Q^2$ since the ladder truncation of the BSE kernel
does not generate a width for any meson bound state.  For example, one 
would have to supplement the ladder BSE kernel with the intermediate 
state $\pi\pi$ production mechanism to produce a width for the 
corresponding $\rho$-meson resonance in the vertex beyond the
threshold for pion production, $Q^2 < -4\,m_\pi^2$ in the timelike
region.

\TABLE[ht]{
\caption{\label{g_fit} Coupling constants for vector meson strong 
decays extracted from timelike charge form factors.}
\begin{tabular}{c|cc}
                &  experiment       & theory \\ \hline
$g_{\rho\pi\pi}$& $6.02 \pm .05$        &  5.4     \\
$g_{\phi KK}$   & $4.64 \pm .14$        &  4.3
\end{tabular}
}

Our results shown in Fig.~\ref{timelikeFF} indeed indicate poles at
\mbox{$Q^2=-0.55~{\rm GeV}^2$} in the two form factors involving
$\bar{u}\gamma u$ coupling and a pole at $Q^2= -1.1~{\rm GeV}^2$ in
the form factor involving $\bar{s}\gamma s$ coupling.  These are the
$\rho$ and $\phi$ respectively and the coupling constants
$g_{\rho\pi\pi}$ and $g_{\phi K K}$ can be extracted from a fit of
Eq.~(\ref{rhopole}) to the calculated timelike charge form factors.
The results are given in Table~\ref{g_fit}, and agree reasonably well
with experiment.  The theoretical or
experimental width is a significantly greater fraction of the
$\rho$ mass than it is of the $\phi$ mass.  Thus the omission of meson
loops in the BSE kernel can be expected to lead to greater error in
the $\rho$ BS amplitude than in the $\phi$ BS amplitude.  For this
reason we speculate that this is also the likely explanation of why the
present impulse approximation for $g_{\rho\pi\pi}$ is further from
experiment than the corresponding result for $g_{\phi KK}$.  }

\subsection{The $\gamma^\star \pi \rho$ and
$\gamma^\star \pi \gamma$ form factors}
{\sloppy
The \mbox{$\pi^0 \to \gamma \gamma$} decay rate is dictated by the
axial anomaly as prescribed by electromagnetic gauge invariance and
chiral symmetry.  The resulting invariant amplitude, for photon 
helicities $\lambda_1$ and $\lambda_2$, is
\begin{eqnarray}
{\cal M}^{\lambda_1 \lambda_2}
   &=& \frac{2\,i\, \alpha_{\rm em} \,g_{\pi \gamma \gamma} }
       {\pi \, \tilde{f}_\pi } 
\nonumber\\ && {} \times
	\epsilon_{\mu \nu \alpha \beta } \, 
	\epsilon^{(\lambda_1)}_\mu \epsilon^{(\lambda_2)}_\nu 
	k_{1\alpha}k_{2\beta}  \; ,
\label{invM}
\end{eqnarray}
where $\epsilon^{(\lambda_i)}_\mu$ and $k_i$ are the polarization
vector and momentum of the $i^{th}$ photon, and $\alpha_{\rm em}$ is
the electromagnetic fine structure constant $e^2/4\pi$.  As indicated
in Eq.~(\ref{invM}), this result is conventionally expressed in terms
of \mbox{$\tilde{f}_\pi = f_\pi /\sqrt{2}$}, the pion decay constant
in the convention where its value is $92$ MeV, rather than the
convention \mbox{$f_\pi=131$} MeV used throughout this present work.
The coupling constant $g_{\pi\gamma\gamma}$ appearing in
Eq.~(\ref{invM}) has the chiral limit value
\mbox{$g^{0}_{\pi\gamma\gamma}=1/2$} and the resulting invariant
amplitude for $\pi^0 \gamma \gamma$ coupling saturates the axial
anomaly through the massless pion pole in the divergence of the axial
current as coupled to the electromagnetic field.  The resulting decay
width
\begin{eqnarray}
 \Gamma_{\pi^0 \gamma \gamma} &=&
	\frac{(g_{\pi\gamma\gamma}^0)^2 \, \alpha^2_{\rm em} \, m_\pi^3}
	{16 \pi^3 \tilde{f}_\pi^2} \; ,
\end{eqnarray}
is within 2\% of the experimental width of $7.8~{\rm eV}$.  The 
corrections due to finite pion mass are small.

It is known that these features of the axial anomaly are preserved by
the ladder-rainbow truncation of the DSEs combined with an impulse
approximation for the $\pi^0 \gamma \gamma$ vertex because the
relevant manifestations of electromagnetic gauge invariance and chiral
symmetry are present~\cite{roberts96,MR98}.

It is convenient to separate out the photon polarization vectors so
that we deal with the vertex $\Lambda^{\gamma \pi \gamma}_{\mu \nu}$
defined by \mbox{${\cal M}^{\lambda_1 \lambda_2} = 
 \epsilon^{(\lambda_1)}_\mu \Lambda^{\gamma \pi \gamma}_{\mu \nu} 
 \epsilon^{(\lambda_2)}_\nu $}.
Corresponding to the flavor decomposition 
\mbox{$\pi^0 = (u\bar u - d\bar d)/\sqrt{2}$}, the  impulse approximation
or triangle diagram for the vertex  has the flavor decomposition
\begin{equation}
\label{piggver}
  \Lambda^{\gamma \pi \gamma}_{\mu \nu}(k_1; k_2) =
      	\left(\hat{Q}_u^2 \, \Lambda^u_{\mu \nu}
         - \hat{Q}_d^2 \, \Lambda^d_{\mu \nu} \right)/\sqrt{2} \,,
\end{equation}
where $\hat{Q}_{u/d}$ are the quark charges.  We use isospin symmetry
where \mbox{$\Lambda^u_{\mu \nu} = \Lambda^d_{\mu\nu}$}.  For each of
these vertices there are two permutations of the final state photons
to be included. The resulting expression is\footnote{The prefactor in
Eq.~(\ref{tri_pigg}) is different than that in
Refs.~\cite{Dub98,Maris:Heidelb} because a) the BSE normalization
conventions are different and b) the symmetry factor of $2$ from the
identity of the decay photons is here included in the vertex.}
\begin{eqnarray}
 \Lambda^{\gamma \pi \gamma}_{\mu\nu}(k_1; k_2) &=&
        \case{\sqrt{2}\, N_c}{3} \int^\Lambda_q \!{\rm tr}_{\rm s}\, 
	\big[ S(q) \, \Gamma^\pi(q,k) \, S(k)  
\nonumber \\  &\times&
        i \Gamma^\gamma_\mu(k,p)\, S(p) \,
        i \Gamma^\gamma_\nu(p,q) \big] \;.
\label{tri_pigg}
\end{eqnarray}
Here the momenta $p, k$ are determined in terms of the integration
momentum $q$ and the outgoing external momenta $k_1, k_2$ by momentum
conservation.
%
\FIGURE[ht]{
\epsfig{figure=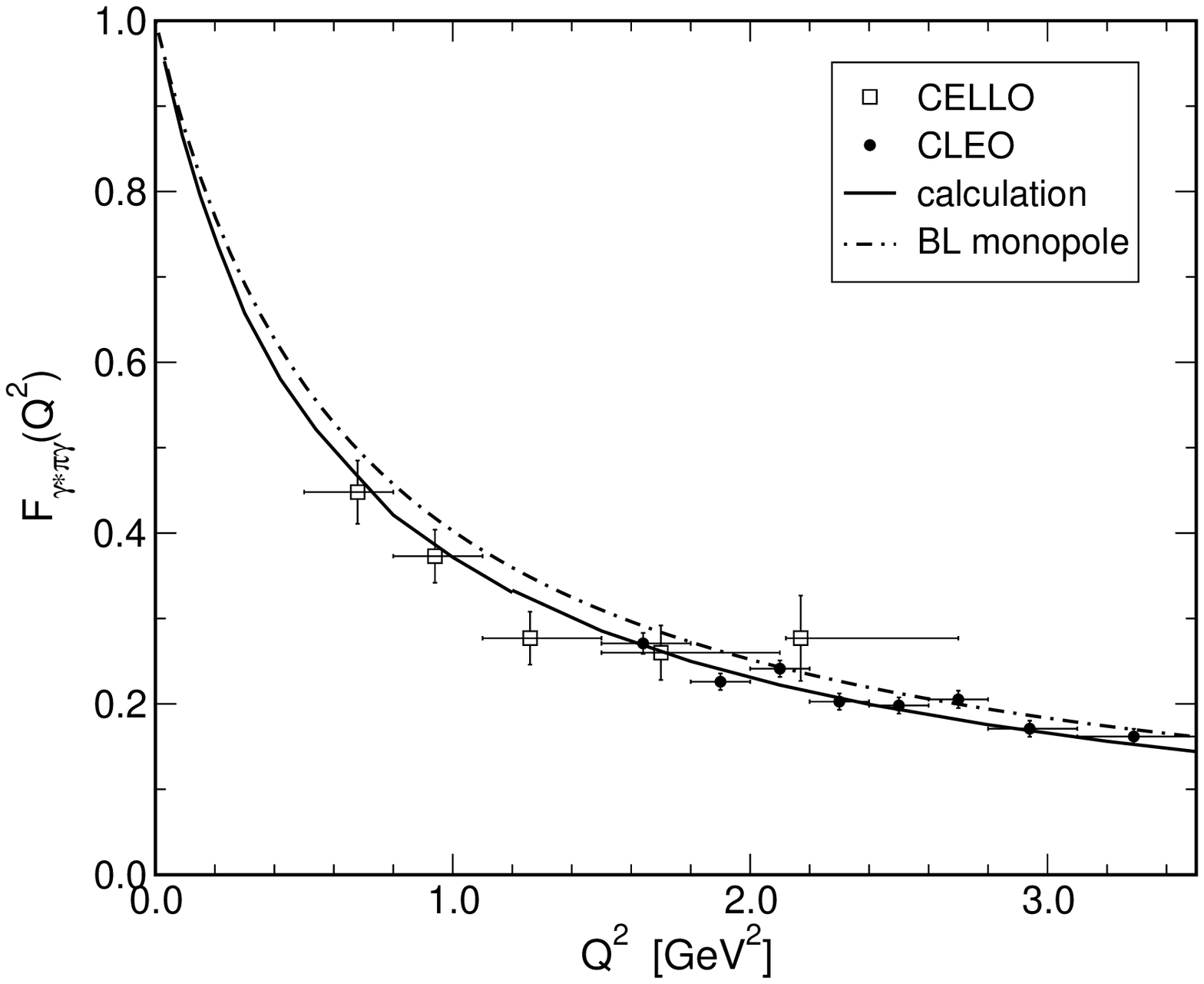,height=5.6cm}
\caption{The $\gamma^\star\,\pi\gamma$ form 
factor~\protect\cite{Maris:Heidelb}, with data from CLEO and
CELLO~\protect\cite{cellocleo}. \label{fig:piggff}}
}
%

To conveniently describe the $\gamma^\star \pi \gamma$ transition and 
define the associated form factor, we replace $k_1$ by the virtual photon 
momentum $Q$ and express $k_2$ as \mbox{$k_2 = P - Q/2$} where the 
incoming pion momentum is \mbox{$P + Q/2$}.   Thus the new independent 
external momenta are $P$ and $Q$ and this facilitates comparisons with 
our descriptions of meson elastic vertices and form factors.  
With one virtual photon, the general form of the vertex is
\begin{eqnarray}
\Lambda^{\gamma\pi\gamma}_{\mu\nu}(P,Q)
   &=& \frac{2\,i\, \alpha_{\rm em} \,g_{\pi \gamma \gamma} }
       {\pi \, \tilde{f}_\pi }
        \,\epsilon_{\mu \nu \alpha \beta }\,P_{\alpha }Q_{\beta }
\nonumber\\ && {} \times
        F_{\gamma^\star\pi\gamma}(Q^2)  \,,
\end{eqnarray}
where the defined form factor satisfies \mbox{$F(0)=1$}.

The present model produces the coupling constant shown in 
Table~\ref{g_vpigam} at the physical pion mass; the chiral limit value
is essentially the same within numerical accuracy.
Our result~\cite{Maris:Heidelb,MTgpg} for the transition form factor
is displayed in Fig.~\ref{fig:piggff}.  The corresponding transition
radius, shown in Table~\ref{g_vpigam}, agrees well with the experimental
value reported by the CELLO collaboration~\cite{cellocleo}.
For the moderate range of $Q^2$ displayed in Fig.~\ref{fig:piggff},
both the data and our DSE model are close to the monopole shape fitted
to the Brodsky--Lepage asymptotic limit~\cite{Lepage:1980fj} 
\mbox{$Q^2 F(Q^2) \sim 4 \pi^2 f_\pi^2$} obtained from pQCD in the 
factorized approximation\footnote{This has been converted to the
convention \mbox{$f_\pi = 131$} MeV used in the present work.}.

For the radiative decay \mbox{$\rho^+ \to \pi^+ \gamma$}, the photon
can radiate from the $u$-quark or the $\bar d$-quark giving 
\begin{eqnarray}
\label{rpgver}
 \Lambda^{\rho^+ \pi^+ \gamma}_{\mu \nu} &=&
        \hat{Q}_u \, \Lambda^{u\bar{d},u}_{\mu \nu}
        + \hat{Q}_{\bar{d}} \, \Lambda^{u\bar{d},\bar{d}}_{\mu \nu} \,,
\end{eqnarray}
where $\Lambda^{u\bar{d},q}_{\mu \nu}$ is the vertex having the
indicated quark flavor labelling and containing no charge or flavor
weights associated with external bosons.  With both the $\rho^0$ and
$\pi^0$ given by $(u\bar u - d\bar d)/\sqrt{2}$, the radiative decay
\mbox{$\rho^0 \to \pi^0 \gamma$} can be expressed as
\begin{eqnarray}
\label{r0p0gver}
 \Lambda^{\rho^0 \pi^0 \gamma}_{\mu \nu} &=&
        \hat{Q}_u \, \Lambda^{u\bar{u},u}_{\mu \nu}
        + \hat{Q}_{d} \, \Lambda^{d\bar{d},d}_{\mu \nu} \,,
\end{eqnarray}
where the radiative contribution from both quark and antiquark of the 
same flavor have been combined.  In the isospin symmetric limit, we have
$\Lambda^{u\bar{u},u}_{\mu \nu} = \Lambda^{d\bar{d},d}_{\mu \nu} =
\Lambda^{u\bar{d},u}_{\mu \nu} = -\Lambda^{u\bar{d},\bar{d}}_{\mu\nu}$,
and thus $\rho^\pm$ and $\rho^0$ have identical $\pi \gamma$ radiative
decays at this level.   In impulse approximation the physical vertex 
is~\cite{Dub98}
\begin{eqnarray}
\lefteqn{\Lambda^{\rho \pi \gamma}_{\mu\nu}(P,Q) =
        \case{N_c}{3} \int^\Lambda_q \!{\rm tr}_{\rm s}\, \big[
        S(q) \, \Gamma^\pi(q,q_+) \, S(q_+)  }
\nonumber \\ && {\;\;\;\;} \times
        i \Gamma^\gamma_\mu(q_+,q_-)\, S(q_-) \,
        \bar\Gamma^{\rho}_\nu(q_-,q) \big] \;,
        \;\;\;\;\;\;\;\;\;\;\;\;\;\;\;\;
\label{tri_gpr}
\end{eqnarray}
where $Q$ is the photon momentum and the $\rho$ momentum is
represented by \mbox{$P - Q/2$}.  This is completely analogous to
Eq.~(\ref{tri_pigg}) for the $\gamma^\star \pi \gamma$ vertex if the
on-shell quark-photon vertex is replaced by the $\rho$ BSA.  In fact,
if that photon momentum in Eq.~(\ref{tri_pigg}) is continued into the
timelike region, an analysis of the $\rho^0$ pole contribution in the
manner discussed in Sec.~\ref{sectimelike} will identify
Eq.~(\ref{tri_gpr}) as the $\gamma^\star\pi\rho$ vertex.
 
The \mbox{$\rho \to \pi \gamma$} coupling constant and the
$\gamma^\star\pi\rho$ form factor are identified from the $\rho \pi
\gamma$ vertex according to
\begin{equation}
\label{lam}
 \Lambda^{\rho \pi \gamma}_{\mu\nu}(P,Q)
        = \frac{i\,e\, g_{\rho \pi\gamma} }{m_{\rho}}
        \, \epsilon_{\mu \nu \alpha \beta } \, P_{\alpha }Q_{\beta }
        \, F_{\gamma^\star\pi\rho}(Q^2) \,.
\end{equation}
At the photon point we have \mbox{$F(0)=1$} and $g_{\rho \pi\gamma}$
is the conventional coupling constant associated with the radiative
decay width
\begin{equation}
\Gamma_{\rho \to \pi \gamma} = 
	\frac{\alpha_{\rm em} \;g_{\rho \pi \gamma}^2}{24}
        \; m_\rho \left(1 - \frac{m_\pi^2}{m_\rho^2} \right)^3 \;.
\label{rhoradwidth}
\end{equation}

The radiative decay \mbox{$\omega \to \pi^0 \gamma$} is given by
\begin{eqnarray}
\label{opgver}
\Lambda^{\omega \pi \gamma}_{\mu \nu}(k_\pi, k_\gamma) &=&
                \hat{Q}_u \, \Lambda^{u\bar{u},u}_{\mu \nu}
                - \hat{Q}_d \, \Lambda^{d\bar{d},d}_{\mu \nu} \,,
\end{eqnarray}
where, compared to Eq.~(\ref{r0p0gver}), the change in phase of the
second term here comes from the phase of the $d \bar d$ component of
the $\omega$. With isospin symmetry, we have \mbox{$\Lambda^{\omega
\pi \gamma}_{\mu \nu} =$} \mbox{$\Lambda^{u\bar{u},u}_{\mu \nu}$} and
hence \mbox{$g_{\omega \pi\gamma}/m_\omega = 
3\,g_{\rho\pi\gamma}/m_\rho$}.

\TABLE[ht]{
\caption{\label{g_vpigam} Coupling constant, interaction radius and 
width for the \mbox{$\pi^0 \to \gamma \gamma$} decay along with coupling 
constants and widths for the indicated  vector meson radiative decays. }
\begin{tabular}{c|cc}
             		&  experiment      	& theory 	\\ \hline
$g_{\gamma\pi\gamma}$   & $0.501 \pm .018$      &  0.50  	\\
$r^2_{\gamma\pi\gamma}$ & $0.42 \pm .04~$fm$^2$ &  0.39~fm$^2$  \\
$\Gamma_{\pi^0 \to \gamma \gamma}$ & $7.8 \pm 0.6$~eV & 7.8~eV 	\\ \hline
$g_{\rho\pi\gamma}/m_\rho$ & $0.74 \pm .05$~GeV$^{-1}$ & 0.69~GeV$^{-1}$ \\
$\Gamma_{\rho^+ \to \pi^+ \gamma}$ & $ 68 \pm 7$~keV  & 53~keV 	\\
$\Gamma_{\rho^0 \to \pi^0 \gamma}$ & $102 \pm 25$~keV & 53~keV  \\ \hline
$g_{\omega\pi\gamma}/m_\omega$ & $2.32 \pm .08$~GeV$^{-1}$ & 2.07~GeV$^{-1}$\\
$\Gamma_{\omega \to \pi^0 \gamma}$ & $717 \pm 43$~keV & 479~keV     
\end{tabular}
}
As Eq.~(\ref{lam}) shows, it is $g/m_V$ that is the natural outcome of
the theory for the vector radiative decays, and it is this combination
that we report in Table~\ref{g_vpigam}.  We also give the
corresponding decay widths.  Except for the difference between the
decay of the neutral and charged states of the $\rho$, which is beyond
the reach of the isospin symmetric impulse approximation, the
agreement between theory and experiment is within 10\%.  This is
consistent with the other $\pi$ and $\rho$ observables obtained from
the same model.

The $\gamma^\star \pi\rho$ form factor plays a role in the
interpretation of electron scattering data from light nuclei, because
the isoscalar meson-exchange current contributes significantly to
these processes.  In particular our understanding of the deuteron EM
structure functions for \mbox{$Q^2 \approx 2-6~{\rm GeV}^2$} requires
knowledge of this form factor~\cite{VODG95}.  An initial exploratory
study~\cite{T96} of the $\gamma^\star \pi\rho$ vertex within the
present framework, but employing phenomenology to a much greater
extent, produced a very soft result for the form factor and this is
shown in Fig.~\ref{fig:gprQ2FF} as the Mitchell--Tandy curve.  It was
found~\cite{VODG95} that the resulting meson exchange current
contribution provided a very good description of the elastic deuteron
electromagnetic form factors $A(Q^2)$ and $B(Q^2)$ in the range
\mbox{$2-6~{\rm GeV}^2$} where such effects are important.

\FIGURE[ht]{
\epsfig{figure=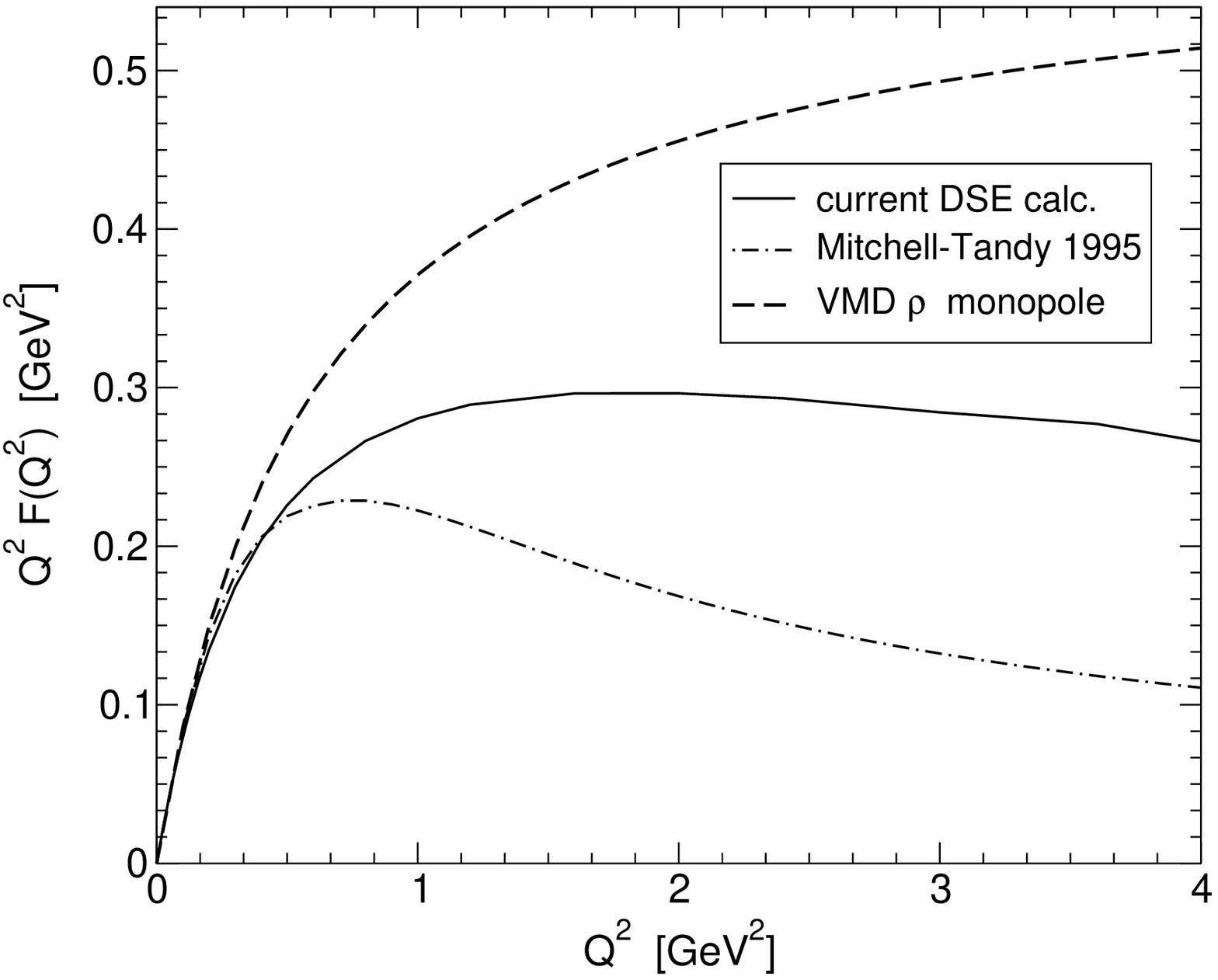,height=5.6cm}
\caption{The $\gamma^\star \pi \rho$ form factor. \label{fig:gprQ2FF} }
}

The model under consideration here, based on the DSEs of QCD, has no
free parameter other than the two set by $f_\pi$ and $\langle q
\bar{q} \rangle$ as described earlier; the amount of phenomenology is
significantly less than the earlier Mitchell--Tandy result~\cite{T96}.
The present work produces a form factor that is much softer than what
is inferred from vector meson dominance (VMD) but obviously harder
than the Mitchell--Tandy result~\cite{T96}, as can be seen in
Fig.~\ref{fig:gprQ2FF}.  We expect the present impulse approximation
results for the form factor to be as reliable as the pion and kaon
charge form factors from this model~\cite{MTpiK00}.  The influence of
these recent results upon the analysis of electron scattering from the
deuteron remains to be determined.  In this regard, some estimation of
the likely uncertainty from the virtuality of the $\pi$ and $\rho$
within the assumed meson exchange current mechanism needs to be made.
}

\section{Extension to finite temperature}
\label{secextfinT}
{\sloppy 
The DSE approach can also be used to study QCD at finite temperature.
This is done most easily via the Matsubara formalism where the energy
components of the 4-momenta become discrete.  Thus the quark
propagator becomes a function of \mbox{$p_n= (\vec{p}, \omega_n)$},
where $\vec{p}$ is a three-vector, and \mbox{$\omega_n=$} 
\mbox{$(2n+1) \pi T$} is the fermion Matsubara frequency.  The most
general decomposition of the dressed renormalised quark propagator 
now becomes
\begin{eqnarray}
 S^{-1}(p_n) & = & i\vec{\gamma}\cdot \vec{p} \,A(p_n)
        + i\gamma_4\,\omega_n \,C(p_n) + B(p_n) \,,
\nonumber \\
\end{eqnarray}
which satisfies the DSE
\begin{eqnarray}
 S^{-1}(p_n) & = & Z_2^A \,i\vec{\gamma}\cdot \vec{p}
        + Z_2^C \, i\gamma_4\,\omega_n
        + Z_4 \,m(\mu)
\nonumber  \\
        & & {} + \Sigma(p_n)\,,
\label{qDSE}
\end{eqnarray}
with quark self-energy given by
\begin{eqnarray}
 \Sigma(p_n) & =& \case{4}{3} \int_{q,l}^\Lambda
        g^2\,D_{\mu\nu}(k_{n-l})\gamma_\mu S(q_l)\Gamma_\nu \, ,
        \;\;\;\;\;\;\;\;
\label{regself}
\end{eqnarray}
where we have employed the notation
\begin{equation}
 \int_{q,l}^\Lambda = T\sum_{l=-\infty}^\infty \!
                \int^\Lambda \! \frac{d^3q}{(2\pi)^3}\,.
\end{equation}
Note that the argument of the gluon propagator,
\mbox{$k_{n-l}=p_n-q_l$}, corresponds to an even (bosonic) Matsubara
frequency.  For renormalization of the propagator we require
\begin{equation}
\label{subren}
 S^{-1}(p_0)\bigg|_{p^2+\omega_0^2=\mu^2} =
        i\vec{\gamma}\cdot \vec{p} + i\gamma_4\,\omega_0 + m(\mu)\;.
\end{equation}
As before, we use the rainbow trunctation for the vertex,
$\Gamma_\nu=\gamma_\nu$, in combination with the ladder truncation for
the BSE kernel $K$. }

\subsection{Chiral symmetry restoration}
{\sloppy 

Chiral symmetry restoration has been studied at both zero and
finite-$(T,\mu)$ in a class of confining DSE models similar to the one
discussed in Secs.~\ref{secladderrainbow} and
\ref{secff}~\cite{bastirev}.  Here, two models are considered in which
the long-range part of the interaction is an integrable infrared
singularity~\cite{mn83} motivated by $T=0$ studies~\cite{pennington}
of the gluon DSE.  Both models can be described by~\cite{hmr98T}
\begin{eqnarray}
g^2 D_{\mu\nu}(k_m) &=&
        P_{\mu\nu}^L {\cal D}(k_m;m_g) +
        P_{\mu\nu}^T {\cal D}(k_m;0) \,,
\nonumber\\
\label{delta}
 {\cal D}(k_m;m_g) &=&
        2\pi^2 D\,\case{2\pi}{T}\delta_{0\,j} \,\delta^3(\vec{k})
\nonumber \\ && {}  + {\cal D}_{\rm M}(k_m^2+m^2_g)\,,
\end{eqnarray}
where $P_{44}^T=P_{4i}^T=0$, $P_{ij}^T=\delta_{ij}-k_i k_j/k^2$,
$P_{\mu\nu}^L=\delta_{\mu\nu}-k_\mu k_\nu/k_m^2-P_{\mu\nu}^T$, and
$m_g$ is a Debye mass.  The strength parameter $D$ is fixed by a fit
to $m_\pi$ and $f_\pi$ at $T=0$.  The models differ in the details of
${\cal D}_{\rm M}$.

It is anticipated that the restoration of chiral symmetry, which
accompanies the formation of a quark-gluon plasma at nonzero
temperature $T$, is a second-order phase transition in QCD with 2
light flavors.  Such transitions are characterised by two critical
exponents: $(\beta,\delta)$, which describe the response of the chiral
order parameters, ${\cal X}$, to changes in $T$ and in the
current-quark mass, $m$.  Denoting the critical temperature by $T_c$,
and introducing the reduced-temperature $t:= T/T_c-1$ and reduced mass
$h:=m/T$, then
\begin{eqnarray}
\label{aa} {\cal X} \propto (-t)^\beta\,,\;\;\; && t\to 0^-\,,\;h=0\,,\\
\label{ab} {\cal X} \propto h^{1/\delta}\,,\;\;\;&& h\to 0^+\,,\;t=0\,.
\end{eqnarray}
Calculating the critical exponents is an important goal because of the
notion of universality, which states that their values depend only on
the symmetries and dimensions, but not on the microscopic details of
the theory.

One order parameter for dynamical chiral symmetry breaking is the
quark condensate~\cite{MR97} $\langle \bar q q\rangle_\mu^0$.  There
are other, equivalent order parameters such as
\begin{equation}
{\cal X}:= B(p=0,\omega_0), \; \; \;
{\cal X}_C:= \frac{B(p=0,\omega_0)}{C(p=0,\omega_0)}.
\end{equation}
They should be equivalent, and the onset of that equivalence is a good
way to determine the domain of reduced mass and reduced temperature in
which the scaling behavior and critical exponents become manifest.  In
Ref.~\cite{hmr98T} it has been verified numerically that in the chiral
limit ($m=0$) and for $t \sim 0$: $f_\pi \propto \langle\bar q
q\rangle \propto {\cal X}(t,0)$; i.e., that these quantities are all
indeed bona fide order parameters.

The success of the nonlinear $\sigma$-model in describing
long-wavelength pion dynamics underlies a conjecture~\cite{pisarski}
that chiral symmetry restoration at finite $T$ in 2-flavor QCD is in
the same universality class as the 3-dimensional, $N=4$ Heisenberg
magnet ($O(4)$ model), with critical exponents~\cite{ofour}
$\beta^H=0.38 \pm .01$, $\delta^H=4.82 \pm .05$.  This is to be
contrasted with mean-field critical exponents, $\beta^{\rm mf}=0.5$,
$\delta^{\rm mf}=3$. }

\subsection{Infrared dominant model}
\label{secIRdom}
{\sloppy 
A simple model~\cite{mn83} that is considered in recent
work~\cite{hmr98T,brs,mrs,arnea} is the infrared dominant [ID] model
with \mbox{${\cal D}_{\rm M}(s) \equiv 0$} in Eq.~(\ref{delta}), with
the mass-scale \mbox{$D=0.56\,{\rm GeV}^2$} fixed by fitting $\pi$-
and $\rho$-meson masses at $T=0$.  A current-quark mass of $m=12\,{\rm
MeV}$ yields \mbox{$m_\pi=140\,{\rm MeV}$}.  There are two chiral
limit solutions corresponding to two phases: the symmetric
Wigner--Weyl solution in which \mbox{$B_0(p_n)=0$}, and the chirally
broken Nambu--Goldstone solution in which
\begin{eqnarray*}
B_0(p_n) & = &\left\{
\begin{array}{lcl}
\sqrt{2\,D - 4 p_n^2}\,, & &p_n^2 < \case{D}{2}\\
0\,, & & {\rm otherwise}.
\end{array}\right.
\end{eqnarray*}
From Eq.~(\ref{aa}), the critical exponent $\beta$ for chiral
restoration is simply \mbox{$\beta = 1/2$} and $T_c = \sqrt{2D} /2\pi$
$\sim 170~{\rm MeV}$.  With a finite current-quark mass the quark DSE
becomes a fourth-order algebraic equation and it is straightforward to
establish~\cite{arnea} from Eq.~(\ref{ab}) that \mbox{$\delta = 3$}.
Clearly, this model has mean field critical exponents.  The question
is: is this result model-dependent or is it a more general property? }

\subsection{1-loop perturbative QCD for the UV}
\label{secmodelC}
{\sloppy
To provide a more realistic ultraviolet behavior, one can add a
short-range term to the $\delta$-function, corresponding to the
exchange of a perturbative gluon at momenta \mbox{$p^2 \gg
\Lambda_{\rm QCD}^2$}.  In Ref.~\cite{hmr98T}, chiral symmetry
restoration at finite temperature is studied using 
\mbox{${\cal D}_{\rm M}(s)= {\cal G}(s)/s$} in Eq.~(\ref{delta}) 
where ${\cal G}$ is given by Eq.~(\ref{gvk2})\footnote{Here a
combination of a Gaussian function and $\delta$-function is used to
represent the strength of the interaction at small momenta, whereas 
in Sec.~\ref{secff}, we used Eq.~(\ref{gvk2}) without an additional
$\delta$-function.  Hence the results described here use
\mbox{$D=0.78~{\rm GeV}^2$} which is lower than the value
\mbox{$D=0.93~{\rm GeV}^2$} of the previous section.  Qualitatively,
the results are independent of the details of the effective
interaction, as long as it is strong enough to break chiral symmetry
at $T=0$.}.  Again, the parameters are fixed at $T=0$ by requiring a
good fit to $m_{\pi/K}$ and $f_{\pi/K}$~\cite{MR97}.

The Debye mass is \mbox{$m_g^2 = (16/5) \pi^2 T^2$} and two values of
the parameter $\omega$ are used which are phenomenologically
equivalent at \mbox{$T=0$}: $\omega_1=0.6m_t$ and $\omega_2=1.2m_t$.
To fix \mbox{$m_\pi=140$}~MeV, the former requires a renormalization
point invariant current-quark mass of \mbox{$\hat{m}_u= $}
\mbox{$6.6~{\rm MeV}$}, while the latter requires \mbox{$\hat{m}_u= $}
\mbox{$5.7~{\rm MeV}$}.  The critical temperature can be found from a
calculation of the order parameter in the chiral limit,
\mbox{$\hat{m}=0$}, as is shown in Fig.~\ref{fig:beta}.  Once $T_c$ is
known, one can analyse the behavior of the order parameter as a
function of the current mass at $T=T_c$, see Fig.~\ref{fig:delta}.
%
\FIGURE[ht]{
\epsfig{figure=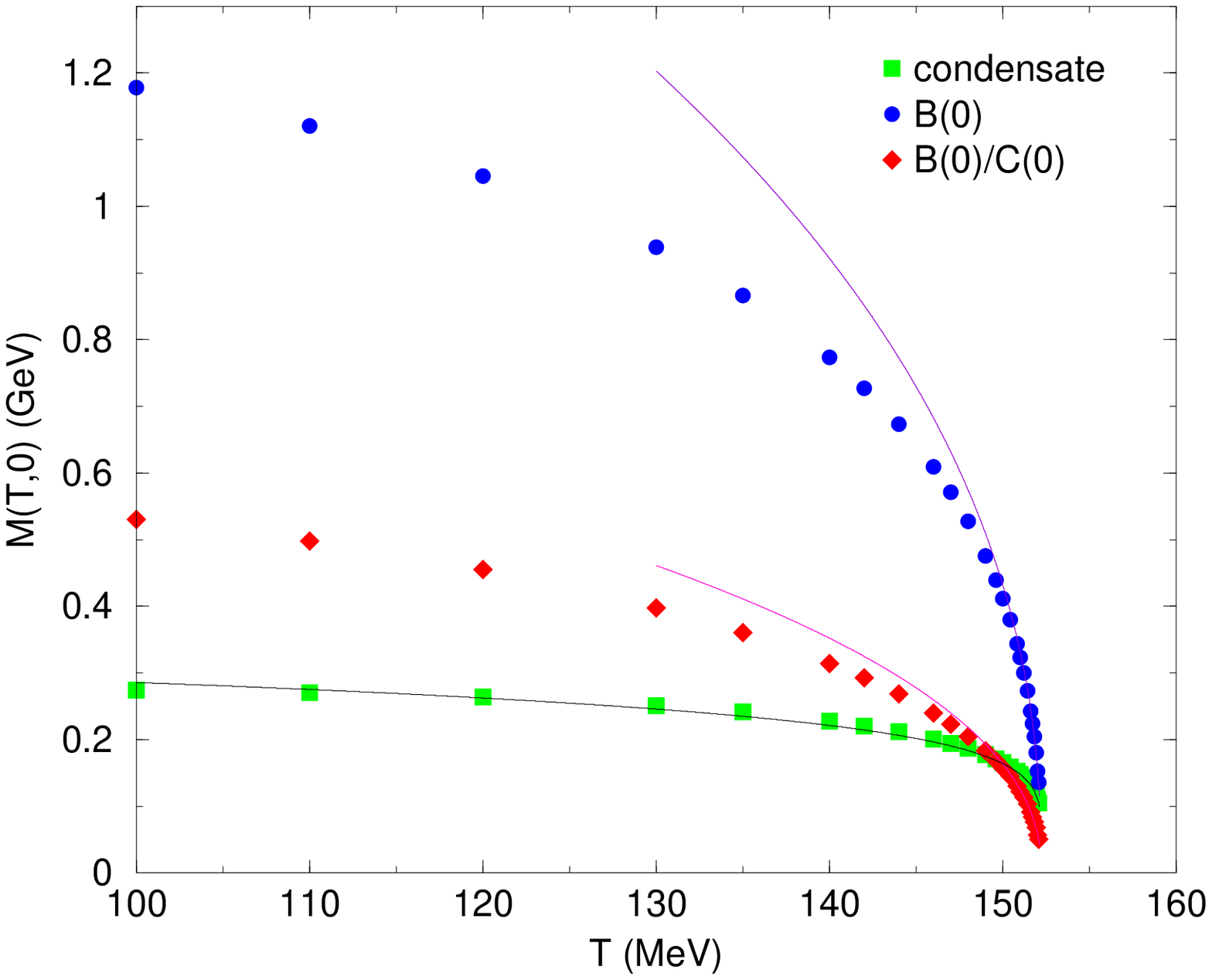,height=5.5cm}
\caption{The analysis of 3 equivalent order parameters for the 
critical exponent $\beta$ within model $\omega_2$. \label{fig:beta} }
}
%
%
\FIGURE[hb]{
\epsfig{figure=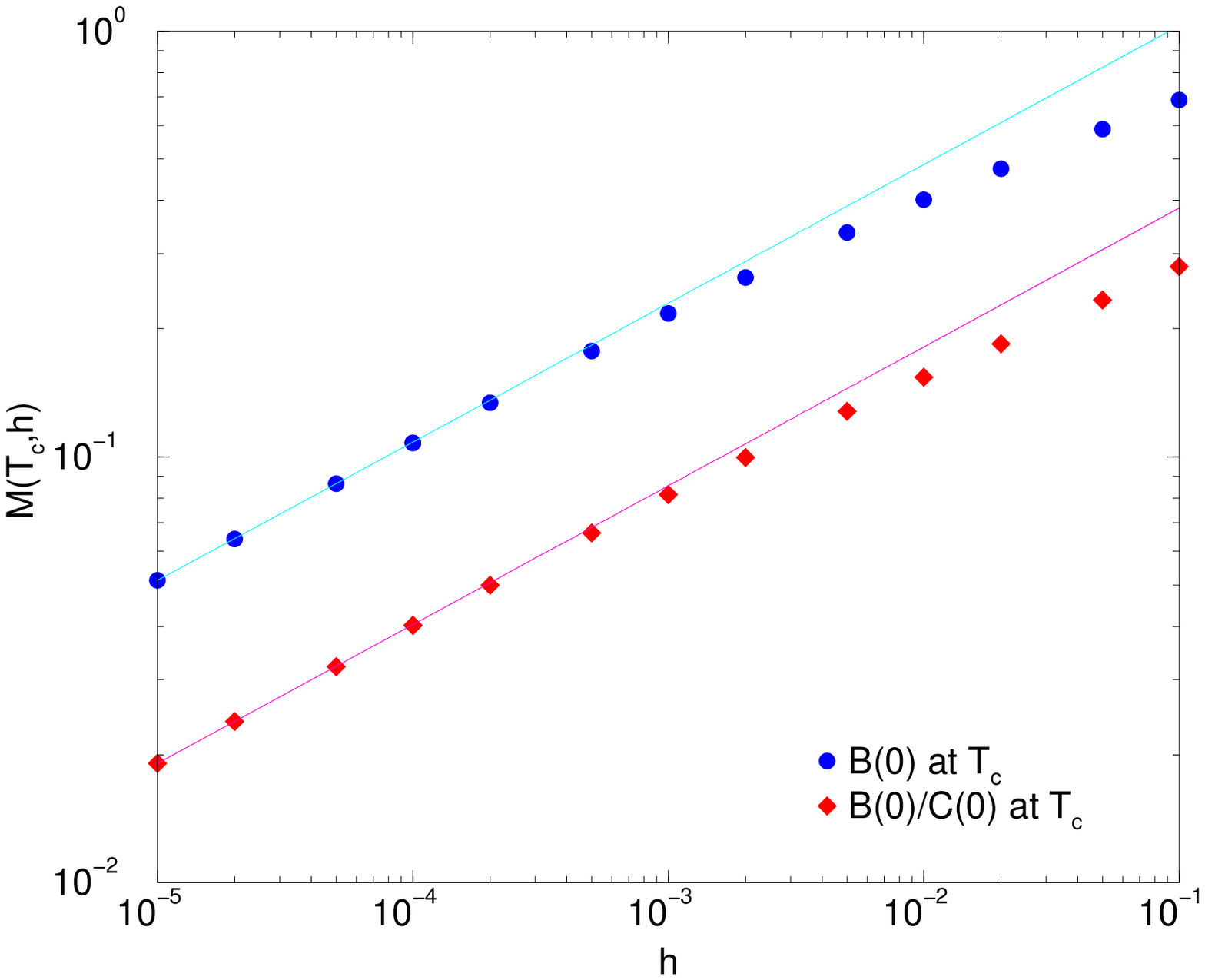,height=5.3cm}
\caption{The analysis of 2 equivalent order parameters for the critical
exponent $\delta$ within model $\omega_2$. \label{fig:delta} } }
%

Although one can in principle extract the critical exponents $\beta$
and $\delta$ from these plots, a much better method is to use the
chiral and thermal susceptibilities~\cite{arnea}.  The critical
exponents obtained from the susceptibilities~\cite{hmr98T} are in good
agreement with a direct calculation of the critical exponents using
Eqs.~(\ref{aa}) and (\ref{ab}), and are given in Table~\ref{expT}.
Note that the scaling relations are only valid for very small
current-quark masses: \mbox{$\log_{10}(\hat{m}/\hat{m}_u) < -5$},
where $\hat{m}_u$ corresponds to the current-quark mass that gives
\mbox{$m_\pi =$} \mbox{$140\,{\rm MeV}$} in this model.  It is only in
this scaling region that the different order parameters become
equivalent.  These results are qualitatively, and for the critical
exponents quantitatively, independent of the parameters in this
model. }
\TABLE[ht]{
\caption{
Critical temperature for chiral symmetry restoration and critical
exponents characterising the second-order transition in the three
illustrative models~\protect{\cite{hmr98T}}.\label{expT}}
\begin{tabular}{c|ccc}
            &   ID      & $\omega_1$ & $\omega_2$ \\ \hline
$T_c$ (MeV) &  169      &   120 & 152   \\ \hline
$\beta$     &$\frac{1}{2}$&0.50 &  0.50 \\
$\delta$    &   3       &  3.00 &  3.02
\end{tabular}
}

\section{Spatial $\bar q q$ modes}
\label{secspatialqq}
At \mbox{$T=0$} the mass-shell condition for a meson as a $\bar q q$
bound state of the BSE is equivalent to the appearance of a pole in
the $\bar q q$ scattering amplitude as a function of $P^2$.  At
$T\neq0$ in the Matsubara formalism, the $O(4)$ symmetry is broken by
the heat bath and one has \mbox{$P \to (\vec{P},\Omega_m)$} where
\mbox{$\Omega_m = 2m \pi T$}.  Bound states and the poles they
generate in propagators may be investigated through polarization
tensors, correlators or Bethe--Salpeter eigenvalues.  This pole
structure is characterized by information at discrete points
$\Omega_m$ on the imaginary energy axis and at a continuum of
3-momenta.  Analytic continuation for construction of real-time
Green's functions (and related propagation properties) has been
well-studied~\cite{LvW87}.  An unambiguous result is obtained by the
requirement that the continuation yield a function that is bounded at
complex infinity and analytic off the real axis~\cite{LvW87}.  One may
search for poles as a function of $\vec{P}^2$ thus identifying the
so-called spatial or screening masses for each Matsubara mode.  These
serve as one particular characterization of the propagator and the
\mbox{$T > 0 $} bound states.  Here we discuss the mass dependence of
the poles in the zeroth Matsubara mode, $\Omega_m=0$.

\subsection{Chiral modes near $T_c$}
\label{secsigT}
{\sloppy 
In the Matsubara formalism, the number of coupled equations in the the
quark DSE, \Eq{gendse}, and the meson BSE, \Eq{homBSE}, scales up with
the number of fermion Matsubara frequencies included.  At low
temperature, the necessary large number of such modes combined with
the loss of $O(4)$ symmetry entails that a straightforward extension
of large scale \mbox{$T=0$} calculations is prohibitive.  Meson $\bar
q q$ modes at \mbox{$T > 0$} have often been studied within the
Nambu--Jona-Lasinio model where the contact nature of the effective
interaction allows decoupling of the Matsubara modes and analytic
summation methods, see for example, Ref.~\cite{FF94} and references
therein.  However the lack of quark confinement in that model leads to
unphysical thresholds for $\bar q q$ dissociation.  To learn about the
fate of meson modes as influenced by deconfinement and chiral
restoration, it is desirable to explore the \mbox{$T \neq 0$} properties
of models that, at \mbox{$T=0$}, confine quarks and provide a
realistic description of the light mesons.

Such a study has recently been carried out~\cite{MRST00} within the
model of Sec.~\ref{secmodelC}.  This work was restricted to
temperatures near and above the transition, \mbox{$T \ge 100\,{\rm
MeV}$}, where a maximum of $10$ Matsubara modes were found to be
sufficient; only pseudoscalar and scalar modes were considered.  In
the chiral limit and at $T=0$, there is dymanical chiral symmetry
breaking and thus one finds massless pseudoscalar Goldstone bosons.
The lowest scalar bound state in this model is an idealized
$\sigma$-meson with $m_\sigma = 0.56\,$GeV.  Such a low-mass scalar is
typical of the rainbow-ladder truncation, although there is some model
sensitivity.  The rainbow-ladder truncation yields degenerate
isoscalar and isovector bound states, and ideal flavor mixing in the
$3$-flavor case; improvements beyond the ladder truncation are
required in order to describe the observed scalar mesons.
Furthermore, in the isoscalar-scalar channel, it will be necessary to
include couplings to the dominant $\pi\pi$
mode~\cite{Pennington:2000hj} because of the large $\sigma$ width.  In
the absence of such corrections, the $\sigma$ properties discussed
below are those of an idealized chiral partner of the $\pi$.

The poles in the 0th Matsubara modes of the scalar and pseudoscalar
vertices, $\Gamma_{\rm S}(p_n,k_n;\vec{P},0)$ and $\Gamma_{\rm
PS}(p_n,k_n;\vec{P},0)$ where $\vec{P}=\vec{p}-\vec{k}$, evolve with
$T$ such that at the critical temperature, one has degenerate,
massless scalar and pseudoscalar bound states~\cite{MRST00}.  Above
$T_c$, these bound states persist, becoming increasingly massive with
increasing $T$.  Furthermore, all but the leading Dirac amplitude
vanish above $T_c$ and the surviving pseudoscalar amplitude $E_\pi$ is
pointwise identical to the surviving scalar one.  These results
indicate that the chiral partners are locally identical above $T_c$,
they do not just have the same mass.  Approaching $T_c$ from below,
the $\sigma$ mass behaves as \mbox{$ m_\sigma \propto
(1-T/T_c)^{\beta}$} within numerical errors.  Note that this is the
same behavior as the chiral condensate and other, equivalent order
parameters for chiral symmetry restoration, such as the decay constant
$f_\pi$, as discussed in the previous section.  For nonzero
current-quark masses chiral symmetry restoration is exhibited as a
crossover rather than a phase transition, see Fig.~\ref{fig:piTln}.
%
\FIGURE[hb]{
\epsfig{figure=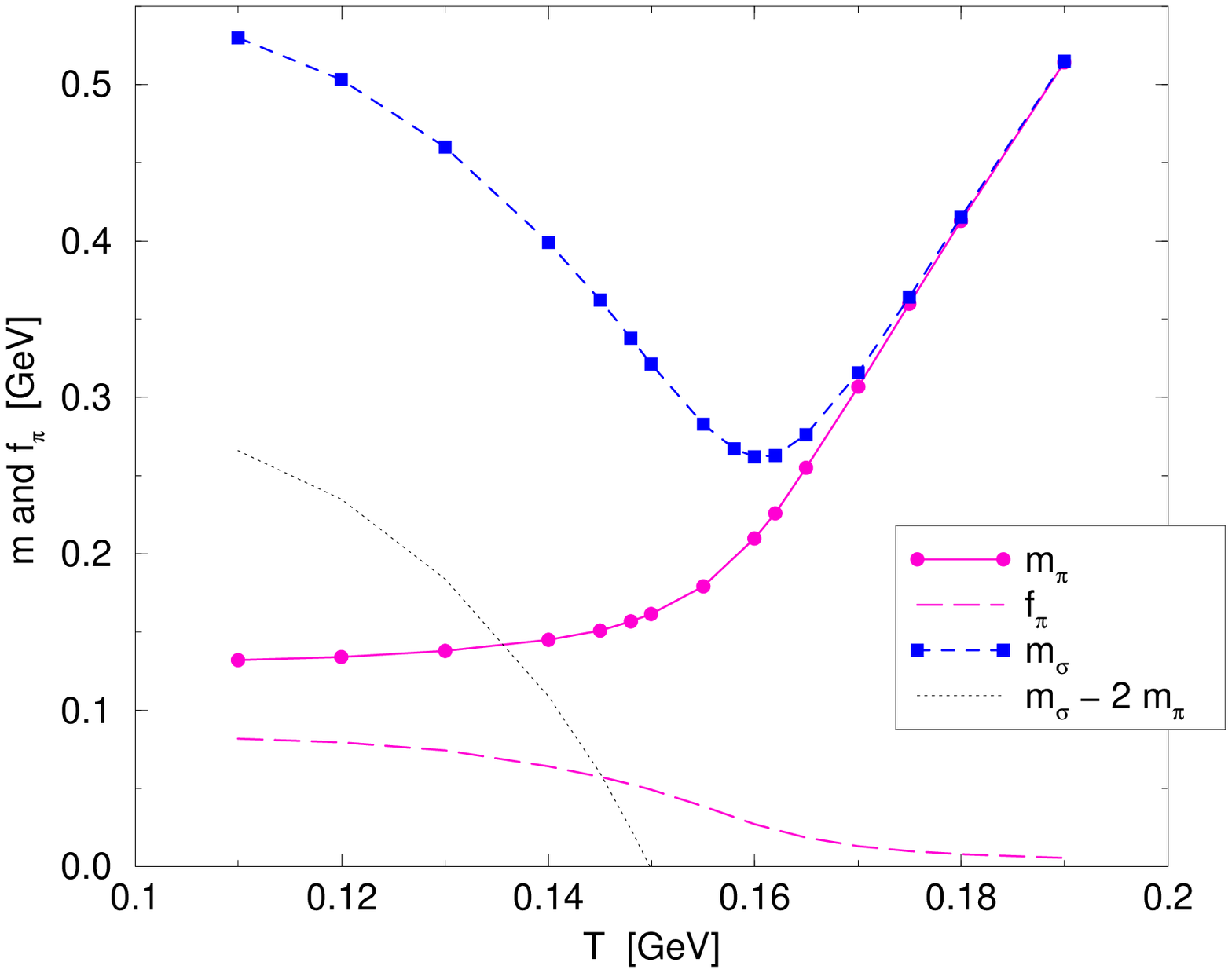,height=5.0cm}
\caption{ 
The $T$-dependence of $m_\pi$ and $f_\pi$ along with the mass 
of the scalar chiral partner of the pion (adapted from 
Ref.~\protect{\cite{MRST00}}).  The model used is $\omega_2$, 
see Table~\protect\ref{expT} and Sec~\protect\ref{secmodelC}.
\label{fig:piTln} }
}
%
The $\sigma$ mass exhibits a dip around the phase transition.  The
scalar and pseudoscalar meson masses become indistinguishable at
\mbox{$T \sim 1.2 \,T_c$}.

With the BSAs obtained from the homogeneous BSE, one can now calculate
the $T$-dependence of other meson properties~\cite{MRST00}.  The decay
of a pion into two photons is calculated in a similar fashion as at
$T=0$ in the previous section.  Using the Ball--Chiu
Ansatz~\cite{BC80} for the photon vertices, one can analytically show
that in the chiral limit the pseudoscalar piece ($E_\pi$) of the pion
BSA~\cite{MR98} saturates the Abelian anomalous contribution to the
divergence of the axial-vector vertex:
\mbox{$g_{\pi\gamma\gamma}=\frac{1}{2}$}.  The tensor structure of
Eq.~(\ref{tri_gpr}) survives at nonzero temperature\footnote
{In this section we use the convention $f_\pi=92\;{\rm MeV}$.}
\begin{equation}
  \Lambda_{i4}(k_1,k_2) = \frac{\alpha_{\rm em}}{\pi}\,
    (\vec{k_1}\times\vec{k_2})_i\,{\cal T}(0)~,
\label{anomT}
\end{equation}
where $k_1$ and $k_2$ are the photon momenta (or rather, the spatial
components of the zeroth Matsubara modes of the photons), and with
\mbox{${\cal T}(0) =$} \mbox{$g_{\pi\gamma\gamma}\,F(0)/f_\pi$}.  
With a suitable extension of the Ball--Chiu Ansatz to nonzero
temperature one can determine the $T$ dependence of the coupling
constant $g_{\pi\gamma\gamma}$, or rather, of
$g_{\pi\gamma\gamma}/f_\pi$ which is the more interesting quantity.
%
\FIGURE[ht]{
\epsfig{figure=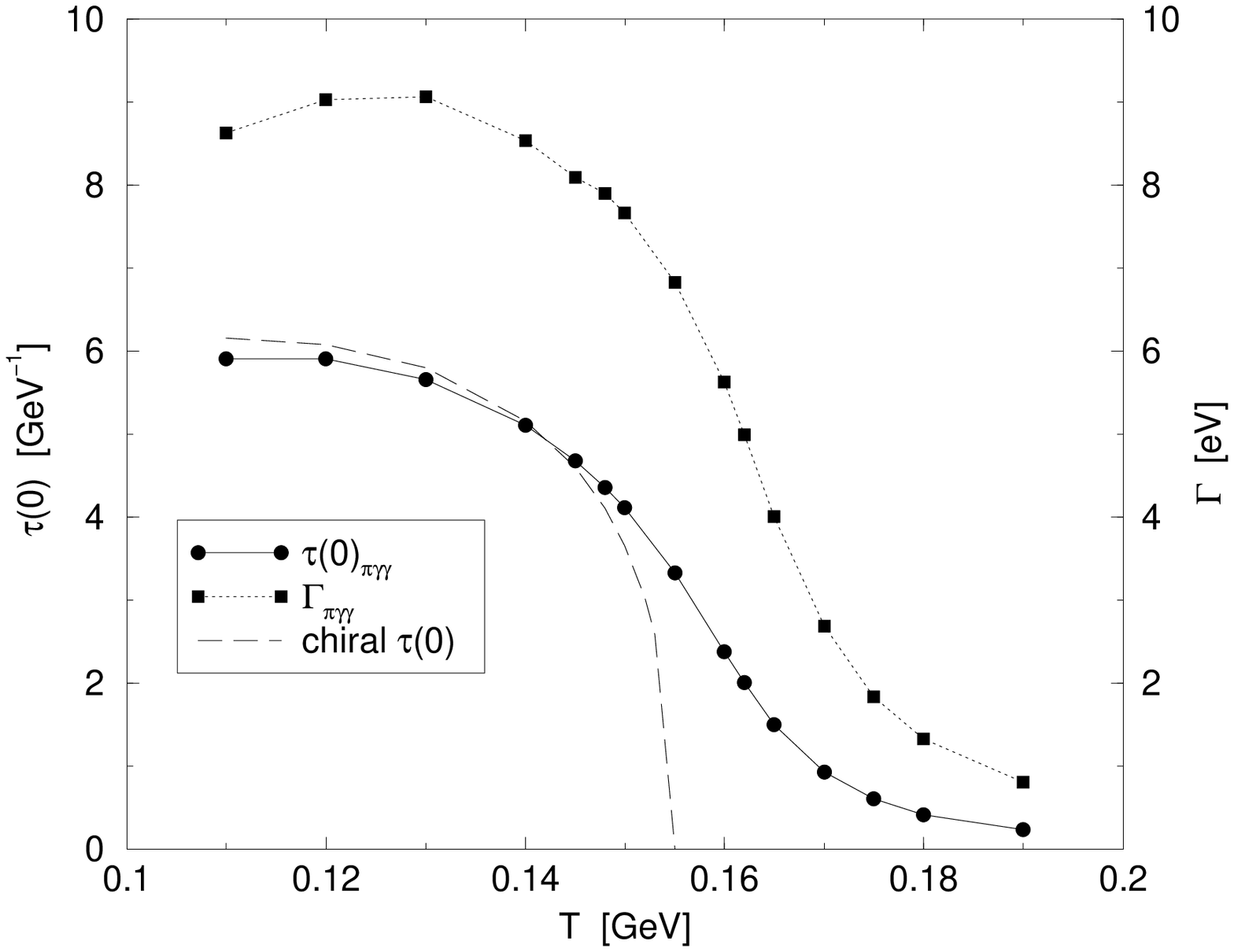,height=5.0cm}
\caption{\label{fig:piwid} $T$-dependence of the coupling ${\cal T}(0)$
and the $\pi^0\to \gamma\gamma$ width (adapted from 
Ref.~\protect{\cite{MRST00}}).}
}
%

The $T$-dependence of the $\pi^0\gamma\gamma$ coupling and the
associated width is depicted in Fig.~\ref{fig:piwid}.  Clear in the
figure is that ${\cal T}(0)$ vanishes at $T_c$; in fact, it vanishes
with the same critical exponents as the other order parameters: $\beta
= 0.5$, indicating mean field behavior~\cite{MRST00}.  An accurate
calculation of the critical exponent is possible because the
Goldberger--Treiman-like relation~\cite{MRT98} $f_\pi^0 E_\pi(p_n) =
B_0(p_n)$ is satisfied for all $T$.  It is therefore not necessary to
solve the BSE for the chiral pion bound state amplitude $E_\pi$.
Thus, in the chiral limit, the coupling to the dominant decay channel
closes for both charged {\it and} neutral pions.  These features were
anticipated in Ref.~\cite{pisarski}.  Further, the calculated 
${\cal T}(0)$ is monotonically decreasing with $T$, supporting the
perturbative analysis to order $(T^2/f_\pi^2)$ in
Ref.~\cite{pisarski2}.  For $\hat m \neq 0$ both the coupling:
$g_{\pi^0\gamma\gamma}/f_\pi$, and the width exhibit the crossover
with a slight enhancement in the width as $T\to T_c$ due to the
increase in $m_\pi$.

The isoscalar-scalar-$\pi\pi$ strong coupling vanishes at $T_c$ in the
chiral limit, which can be traced to $B_{\hbox{\scriptsize{chiral}}}
\to 0$.  For $\hat m \neq 0$, the coupling reflects the crossover.
For the present case of the idealized $\sigma$-meson, there are no
physical consequences because the width for the decay $\sigma \to
\pi\pi$ vanishes just below $T_c$ where the $\sigma$ meson mass falls
below $2 m_\pi$ and the phase space factor vanishes.  
}

\subsection{Separable model}
{\sloppy
For studies at low finite temperature, where continuity with
\mbox{$T=0$} results is to be verified, the required number of
Matsubara modes can be of order $\sim 10^3$.  This is rather
cumbersome with the model discussed above, but can be done if one
utilizes a separable Ansatz~\cite{sepT01,bkt} for the interaction
that can implement the essential qualitative features of DSE/BSE
models at finite temperature.  Separable representations have
previously been found capable of an efficient modeling of the
effective $\bar q q$ interaction in the infrared domain for
\mbox{$T=0$} meson observables~\cite{sep,b+}.  

In Ref.~\cite{sepT01} a simple separable interaction is used, which
confines quarks at \mbox{$T=0$}.  The few parameters are fixed by
$\pi$ and $\rho/\omega$ properties.  The approach is a simplification
of one developed earlier~\cite{b+} that was found to be quite
successful for the light meson spectrum at \mbox{$T=0$}.  The
confining mechanism is an infrared enhancement in the quark-quark
interaction that is strong enough to remove the possibility of a mass
shell pole in the quark propagator for real $p^2$.  In this
implementation, it is particularly transparent that sufficiently high
temperature will necessarily restore a quark mass-shell pole and there
will be a deconfinement transition.  This model also implements low
temperature dynamical chiral symmetry breaking and it preserves the
Goldstone theorem in that the generated $\pi$ is massless in the
chiral limit.  The solutions of the BSE for the $\pi$ and $\rho$ modes
are particularly simple and are used to study the $T$-dependence of
the meson masses and decay constants in the presence of both
deconfinement and chiral restoration mechanisms.

Here we discuss the results obtained~\cite{sepT01} with the rank-2
separable interaction
\begin{eqnarray}
\nonumber
D_{\mu\nu}^{\rm eff}(p-q) &=&
                     \delta_{\mu\nu}\, [D_0~ f_0(p^2)f_0(q^2)  \\
            &+& D_1~ f_1(p^2)(p\cdot q)f_1(q^2)] \; ,
\label{model}
\end{eqnarray}
and a Feynman-like gauge is chosen for phenomenological simplicity.
The choice for the two strength parameters $D_0, D_1$, and
corresponding form factors $f_i(p^2)$ is constrained by consideration
of the resulting solution of the DSE for the quark propagator in the
rainbow approximation, and fitted to reproduce $\pi$ and $\rho$
properties at $T=0$.  In Ref.~\cite{sepT01} a single Gaussian
representation for the $f_i(p^2)$ was used to represent the major
features.  The finite $T$ extension is implemented using the Matsubara
formalism as before, via the discretization of the timelike component
of momenta: $p \rightarrow p_n = (\vec{p}, \omega_n)$.

%
\FIGURE[ht]{
\epsfig{figure=r2ABC.eps,height=6.5cm,angle=-90}
\caption{\label{fig:ABC} 
$T$-dependence of quark self-energy amplitudes at \mbox{$p=0$}
(adapted from Ref.~\protect\cite{sepT01}). }
}
%
The $T$-dependence of the solutions for $A, B$ and $C$ at
\mbox{$\vec{p}^2=0$} is displayed in Fig.~\ref{fig:ABC}.  The results
shown are for the lowest Matsubara mode ($\omega_0=\pi\,T$) which
provides the leading behavior as $T$ is increased.  The chiral
restoration critical temperature $T_c = 121\;{\rm MeV}$ is identified
from the vanishing of the chiral limit amplitude $B_0(0,\omega_0)$ as
shown.  Below $T_c$, $A$ and $C$ are relatively constant, $O(4)$
symmetry is approximately manifest, and the main effect is an almost
constant quark wave function renormalization via $1/C$; the central
feature is a rapidly decreasing mass function\footnote{The rank-1
limit of this model has \mbox{$A=C=1$} for all $T$ and the behavior of
$B(0,\omega_0)$ as function of $T$ is similar to that in \Fig{fig:ABC}
except \mbox{$T_c=$} 146 MeV is obtained~\cite{sepT01}.}.  Above
$T_c$, there remains a significant temperature range where the
self-interaction effects are strong, both $A$ and $C$ are considerably
enhanced above their perturbative values, and the breaking of $O(4)$
symmetry is manifest.  The present model thus captures the qualitative
$T$-dependence observed for the dressed quark propagator in studies of
the quark DSE~\cite{arnea,bbkr}.

The deconfinement transition is characterized by the appearance
of a quark progagator pole for real values of $\vec{p}^2$.
The chiral restoration point $T_c$ and the quark deconfinement point
$T_d$ are generally expected be to identical or nearly
so~\cite{laermann}.  The separable model under consideration here
produces \mbox{$T_d=0.9~T_c = 105\;{\rm MeV}$}.  
}

\subsection{Vector mesons at nonzero $T$}
{\sloppy
Due to the breaking of $0(4)$ invariance, the \mbox{$T=0$} transverse
vector meson splits for $T > 0$ into 3-space longitudinal and
transverse modes.  For the spatial modes characterized by
\mbox{$(\vec{P},\Omega_0=0)$}, the BSAs studied in the separable
model are~\cite{sepT01}
\beq
 \Gamma^{\rho(L)}_\mu(p_n,k_n;\vec{P}) = 
        \delta_{\mu 4} \, \gamma_4 \,
        f_0(q_n^2) F_{\rho(L)}(\vec{P}^2) \; ,
 \label{BSrhoTL}
\eeq
and
\beqar
&&\Gamma^{\rho(T)}_i(p_n,k_n;\vec{P}) =
\nonumber  \\ 
&&\left( \gamma_i - 
        \frac{P_i \vec{P} \cdot \vec{\gamma} }{\vec{P}^2} \right)
                        f_0(q_n^2) F_{\rho(T)}(\vec{P}^2)  \; ,
\label{BSrhoTT}
\eeqar where $q_n = (p_n+k_n)/2$ is the relative momentum.  The
$T$-dependence of the corresponding masses is displayed in
Fig.~\ref{fig:rhomass}.  These modes are almost degenerate and
$T$-independent until about $T_c/2$ where the breaking of $O(4)$
invariance becomes significant.  The qualitative features
\mbox{$M^L_\rho(T) > M^T_\rho(T)$} and \mbox{$M^T_\rho (T) \approx
{\rm constant}$} for \mbox{$T<T_c$} seen here in the context of a
finite range interaction have previously been noted within the
limiting case of the zero momentum range ID model~\cite{mrs}.  This
latter model was not studied for $T>T_c$.  In the separable model
discussed here, the longitudinal mode becomes unstable to $\bar q q$
dissociation at $T \sim $ 180 MeV; the transverse mode continues to be
below the spatial $\bar q q$ threshold for the temperature range
displayed. 
%
\FIGURE[ht]{
\epsfig{figure=r2vmass.eps,height=6.0cm,angle=-90}
\caption{\label{fig:rhomass} $T$-dependence of the longitudinal $\rho$
mass (dot-dashed line) and the transverse $\rho$ mass (solid line).
Also shown is $2 M_\pi$ indicating that the strong decay channel becomes
inaccessible within 20-30 MeV beyond $T_c$
(adapted from Ref.~\protect\cite{sepT01}). }
}
%

Next, one can study the couplings appropriate to \mbox{$\rho^0 \to$}
\mbox{$ e^+\,e^-$} and \mbox{$\rho \to$} \mbox{$ \pi\pi$} for the
meson modes characterized by \mbox{$P=(\vec{P},0)$}~\cite{sepT01}.
The various spatial modes from \mbox{$\Omega_m \neq 0$} are
characterized by masses much greater than that of the lowest mode
considered here.  It is anticipated that this lowest mode
characterizes the qualitative behavior of the physical decay processes
\mbox{$\rho^0 \to e^+\,e^-$} and \mbox{$\rho \to \pi\pi$}. 

The \mbox{$T=0$} expression for the electromagnetic coupling constant
$g_\rho$ has a straightforward extension to \mbox{$T>0$} for a
transverse spatial $\rho^0$ mode~\cite{sepT01}.  Use of the
\mbox{$\Omega_m=0$} solution described above yields
\begin{eqnarray}
\lefteqn{ \frac{{M^T_\rho}^2(T)}{g_\rho(T)} = }
\nonumber \\  && 
        \frac{N_c}{2} \int_{q,n}^\Lambda\! {\rm tr}_{\rm s} 
        \big[ \gamma_i \, S(q_{n+})\, 
        \Gamma_i^{\rho(T)}(q_{n+},q_{n-};\vec{P}) \, S(q_{n-})\big]\,.
\nonumber \\ {}
\label{rhophoton}
\end{eqnarray}
The impulse approximation for the $\rho\pi\pi$
vertex~\cite{pctrev,Dub98}, after extension to \mbox{$T>0$} for
spatial modes characterized by \mbox{$Q=(\vec{Q},0)$} for the $\rho$
and \mbox{$P=(\vec{P},0)$} for the relative $\pi\pi$ momentum, takes
the form
\begin{eqnarray}
\lefteqn{ \Lambda_\nu(P,Q) = P_\nu \, g_{\rho \pi\pi}(T) = }
\nonumber \\  &&
        -2 N_c \int_{q,n}^\Lambda \,{\rm tr}_{\rm s}\, \big[
         S(q)\,\Gamma_\pi(q,q_{n+};-\vec{P}_+)\, S(q_{n+})\,
\nonumber \\ &&
 \times  \Gamma_\nu^{\rho(T)}(q_{n+},q_{n-};\vec{Q})\, 
        S(q_{n-})\,\Gamma_\pi(q_{n-},q;\vec{P}_-) \big]\,,
\nonumber \\ {}
\label{gpp}
\end{eqnarray}
where \mbox{$\vec{P}_\pm = \vec{P} \pm \vec{Q}/2$}.  Use of
\Eq{BSrhoTL} immediately shows that the longitudinal $\rho$ mode
cannot couple to $\pi\pi$.  With summation over enough Matsubara modes
for convergence, both the electromagnetic coupling $g_\rho(T)$, and
the strong coupling $g_{\rho \pi\pi}(T)$, reproduce the independently
determined \mbox{$T=0$} results of this model.  There is very little
$T$-dependence below about $0.8~T_c$ where there is approximate $O(4)$
symmetry as evident in the quark propagator behavior in
Fig.~\ref{fig:ABC}.  Near $T_c$, $g_\rho(T)$ rises and $g_{\rho
\pi\pi}(T)$ decreases.

%
\FIGURE[ht]{
\epsfig{figure=r2rhowid.eps,height=5.0cm,angle=0}
\caption{\label{fig:rhowid} $T$-dependence of the transverse $\rho$ partial
widths due to electromagnetic \mbox{$e^+e^-$} decay (dashed line) and
strong \mbox{$\pi \pi$} decay (solid line)
(adapted from Ref.~\protect\cite{sepT01}). }
}
%
Combined with the relevant phase space, this leads to the
electromagnetic decay width
\begin{eqnarray}
        \Gamma_{\rho^0 \rightarrow e^+\,e^-}(T) &=& 
                \frac{4\pi\,\alpha\,M^T_\rho(T)}{3\;g_\rho^2(T)}~,
\end{eqnarray}
while the corresponding  strong decay width is
\beq
\Gamma_{\rho\rightarrow\pi\pi}(T) = \frac{g_{\rho\pi\pi}^2(T)}{4\pi}
\frac{M_\rho^T(T)}{12}\left[1-\frac{4M_\pi^2(T)}{{M_\rho^T}^2(T)}
                                                      \right]^{3/2}.
\label{rhowidth}
\eeq The $T$-dependence estimated in this way for the decay widths is
due to the response of the quark substructure to the heat bath,
particularly the restoration of chiral symmetry.  The results are
displayed in Fig.~\ref{fig:rhowid}.  The contrast between the behavior
of the electromagnetic and strong widths near and just above $T_c$
should be a more robust finding than the details of the individual
processes.  The strong width decreases rapidly and vanishes just above
$T_c$ while the electromagnetic width remains within 20\% of the
\mbox{$T=0$} value.  Part of the strong decrease of the intrinsic
$\pi\pi$ width of the transverse $\rho$ is due to the decrease in the
coupling constant, however the dominant effect is the $T$-dependence
of the phase space factor in \Eq{rhowidth}.  As displayed in
Fig.~\ref{fig:rhomass}, $2 M_\pi(T)$ rises faster with $T$ than does
$M_\rho^T(T)$ until at \mbox{$T=1.17~T_c$} one has \mbox{$M_\rho^T = 2
M_\pi$}.  Beyond this point, the strong decay \mbox{$\rho^T \to
\pi\pi$} is phase-space blocked.  This suggests that the total
$\rho^T$ width of 151~MeV at $T=0$ decreases by about 50\% near
\mbox{$T=T_c$} and drops sharply to the electromagnetic value of about
6~keV by \mbox{$T=1.17~T_c$}.

One would expect the mass of the pseudoscalar $K$ correlation to rise
with $T$ in a similar fashion to $M_\pi$, while the masses of the
vector $\phi$ and $K^\star$ modes should rise like the $\rho$.  This
suggests that the vector modes $\rho$, $K^\star$ and $\phi$ tend to be
trapped with their relatively long electroweak lifetimes and with
significantly increased masses for a domain of high temperatures above
the transition.  This suggests that within the gas of pions and other
pseudoscalars that dominate the hot hadronic product from heavy-ion
collisions, the role of vector meson correlations in producing the
dilepton spectra could be significantly less than conventional
expectations.

This narrowing of the intrinsic decay width of the vector meson mode
in the heat bath, and also the decrease of the decay width of the
idealized $\sigma$-meson discussed in Sec.~\ref{secsigT}, is a
mechanism that is distinct from the collisional broadening
effect~\cite{RW00} from the many-hadron environment.  The results
discussed here indicate that there is a non-trivial $T$-dependence to
intrinsic coupling constants, such as $g_{\rho \pi \pi}$ and
$g_{\sigma \pi \pi}$, and decay phase space.  The intrinsic effect
tends to significantly decrease the decay width; the hadronic medium
effects have the opposite influence.  An approach that incorporates
both phenomena is clearly called for.

Also note that only the spatial or screening $\bar q q$ masses have been
investigated within this model.  The temporal masses provide a
different characterization of the correlations; lattice simulations
indicate that spatial masses become much larger than the temporal
masses above $T_c$~\cite{TARO99}, whereas a Nambu--Jona-Lasinio model
study~\cite{FF94} found that they are significantly different only in
the range 150 MeV \mbox{$<T<$} 350 MeV.
}

\subsection{Behavior at high $T$}
{\sloppy 
Above the deconfinement and chiral transition temperature, it might be
expected that meson modes should dissolve in favor of a gas of
essentially massless quarks.  However for a significant temperature
range above $T_c$, the spatial $\pi$ and $\rho$ modes studied here
continue to be stable against $\bar q q$ dissociation and do not
dissolve into a free quark gas~\cite{sepT01}.  The results for
$M_\pi(T)$ and $M_\rho^T(T)$ are displayed in \Fig{fig:hiTmass}, along
with the quark dynamical mass function at \mbox{$p=0$}, using the
simpler rank-1 separable interaction.  The masses of both spatial
meson modes approach the asymptotic behavior $2\pi T$ from below.
This asymptotic behavior has been observed in lattice
simulations~\cite{laermann,gocksch} as well.
%
\FIGURE[ht]{
\epsfig{figure=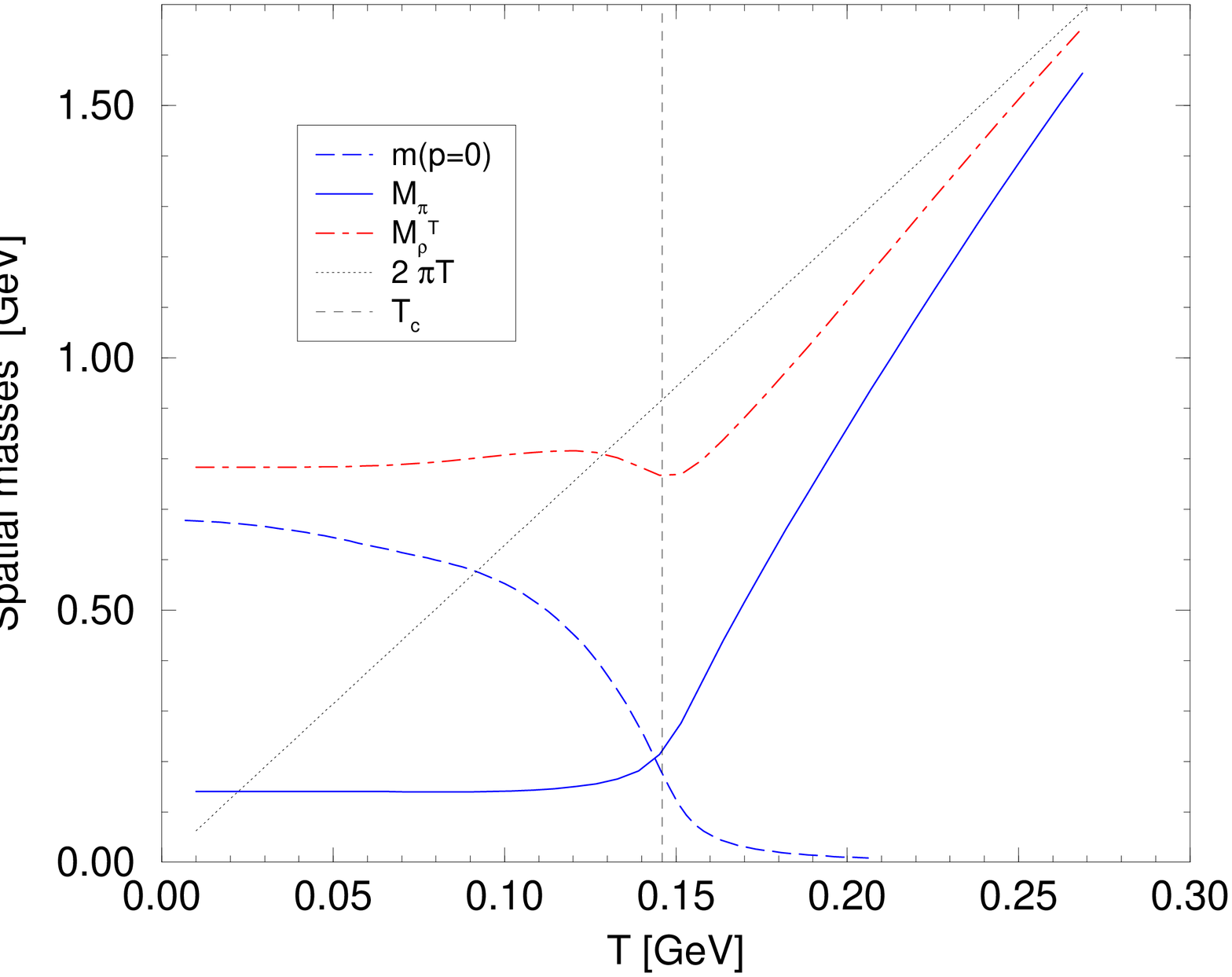,height=6.0cm,angle=-90}
\caption{\label{fig:hiTmass} High $T$ behavior of the spatial masses of
the $\pi$ and transverse $\rho$ modes from the rank-1 separable model
(adapted from Ref.~\protect\cite{sepT01}).}
}
%

The emergence of the $2\pi T$ behavior for masses of $\bar{q} q$
correlations at large $T$ can easily be understood in the rank-1
model, as is shown in Ref.~\cite{sepT01}.  There it is found that for
spatial meson modes \mbox{$M(T) \sim 2\pi T - \Delta(T)$} where
$\Delta$ is a positive mass defect that will typically decrease with
$T$.  Thus the spatial meson mass or screening mass will approach the
thermal mass of a pair of massless fermions from below.

The same qualitative behavior at large $T$ can be seen in a
semi-analytic way by considering the ID model of
Sec.~\ref{secIRdom}.  For example, it is found~\cite{sepT01} that
the leading large $T$ behavior of $M_\rho^T(T)$ is
\beq
{M_\rho^T}^2(T) \to (2\pi T)^2 \, \left( 1 - \frac{\eta}{\pi T}
                          \cdots \right) \; .
\label{asymM}
\eeq
Due to the long range nature of this model, the  asymptotic spatial 
mass defect of the transverse $\rho$ mode actually becomes constant 
\mbox{$\Delta_\rho(T) \to \eta$}.

%
\FIGURE[hb]{
\epsfig{figure=r1mlatt.eps,height=6.5cm,angle=-90}
\caption{\label{fig:latt_cf} 
Spatial masses from the rank-1 separable model (sep) and the ID model
from Ref.~\protect\cite{sepT01} compared to lattice QCD simulations
(latt) taken from Ref.~\protect\cite{gocksch}.}  }
%
In \Fig{fig:latt_cf} the $\pi$ and transverse $\rho$ spatial masses at
\mbox{$T>T_c$} from both the rank-1 separable model and the ID model
are presented in comparison with lattice QCD simulations of spatial
screening masses~\cite{gocksch}.  The solid horizontal line marks
$2\pi$ while the lower horizontal dot-dashed line represents the
lattice free limit corrected~\cite{gocksch} for the lattice time
extent $N_t$.  From \Fig{fig:latt_cf} it is evident that for the $\pi$
at \mbox{$T>T_c$}, the spatial or screening mass defect
\mbox{$\Delta=2\pi T-M(T)$} is decreasing more rapidly above $T_c$ in
the rank-1 separable model than is evident from the lattice
simulations.  This is consistent with the exponential behavior of the
employed form factors.  It is also consistent with the absence of
quark vector self-energy amplitudes $A(p_n)-1$ and $C(p_n)-1$ through
which interactions can be quite persistent in the asymptotic region.
This can be demonstrated within the chiral limit ID model where those
amplitudes are strong and indeed have the power law fall-off. As seen
in \Fig{fig:latt_cf}, the resulting mass defect for both $\pi$ and
$\rho$ is in fact too strong when compared with the lattice QCD
simulations.  This persistent self-interaction well above $T_c$, which
slows the approach to free behavior such as Stefan--Boltzmann
thermodynamics~\cite{brs}, may well be what is signaled by the lattice
QCD data in \Fig{fig:latt_cf}. }

\section{Summary}
\label{secsum}
{\sloppy 
Within a Poincar\'e-covariant approach to modeling QCD based on the
DSE framework, the description of the ground state pseudoscalar and
vector mesons has been reviewed at both zero temperature and finite
temperature.  The only approximation made is a self-consistent
truncation of the set of DSEs, which respects the relevant vector and
axial-vector WTI.  In particular we choose the truncation wherein the
employed quark propagators, the meson BSAs, and the quark-photon
vertex are solutions of their DSEs in rainbow-ladder approximation.
The two parameters are fixed previously by fitting the chiral
condensate, $m_{\pi/K}$ and $f_{\pi}$.  We include all relevant Dirac
amplitudes for the BSAs and their full dependence upon the angular
variable $k\cdot P$.

The obtained pion and kaon electromagnetic form factors are in good
agreement with the available data over the entire $Q^2$ range
considered, and the calculated charge radii are within the bounds of
the experimental values.  The electromagnetic current is explicitly
conserved in this approach, and there is no fine-tuning needed to
obtain \mbox{$F_\pi(0)$} \mbox{$=$} \mbox{$ 1 = $} \mbox{$F_{K^+}(0)$}
and $F_{K^0}(0)=0$.  We also demonstrate explicitly that our results
are (within numerical accuracy) independent of the momentum
partitioning of the BSAs.  The timelike behavior of the charge form
factors exhibits the vector meson production resonances and this
allows the associated vector meson strong decay constants to be
extracted.  The confining property of the quark self-energy
dynamically produced in this work prevents the appearance of spurious
$\bar{q}q$ production effects.  We also discuss transition form
factors for the processes \mbox{$\gamma^\star \rho \to \pi$} and
\mbox{$\gamma^\star \pi \to \gamma$} as well as the coupling constants
for the associated physical radiative decay of the $\rho$ and the
electromagnetic decay of $\pi^0$ as dominated by the axial anomaly.

We discuss the finite temperature extension of the approach as
implemented by the Matsubara method.  Both the chiral restoration and
the deconfinement transitions are identified and a study of the
critical exponents for the chiral transition is reviewed.  Several
versions of the model are employed to gain complementary viewpoints on
the spatial $\bar{q}q$ correlative modes with (pseudo-)scalar and
vector meson quantum numbers.  The corresponding masses $M_\pi(T)$,
$M_\rho^T(T)$ and $M_\rho^L(T)$ are found to be almost $T$-independent
below $T_c$ followed by a strong increase.  This behavior is
characteristic of lattice QCD simulations~\cite{laermann,gocksch} and
DSE studies~\cite{MRST00,sepT01,mrs}.  The scalar mass $M_\sigma(T)$,
an idealized chiral partner of the pion, decreases with $T$ until it
becomes equal to the pion mass at $T_c$; above the critical
temperature for chiral symmetry restoration the scalar and
pseudoscalar modes are degenerate.  Once the scalar mass drops below
twice the pion mass, the decay width associated with the decay $\sigma
\to \pi\pi$ drops to zero, because of phase-space
blocking~\cite{MRST00,scalar}.  Also the strong decay \mbox{$\rho \to
\pi \pi$} decreases with temperature, and drops to zero at about
25~MeV above $T_c$ due a phase-space blocking effect.

One should keep in mind that the results presented here follow from
the response of the quark gluon content of the mesons to the heat
bath.  A different phenomenon is the coupling of the meson modes to
the many-hadron environment which introduces collisional
broadening~\cite{RW00}.  In order to study this effect, one has to go
beyond rainbow-ladder truncation, and incorporate meson loop
corrections.  Also at zero temperature such corrections need to be
studied in more detail: pion and kaon loops give rise to a nonzero
width for the vector mesons.  This would also allow for more realistic
calculations of electromagnetic form factors in the timelike region,
near and beyond the vector meson resonances.  However, it is unclear
how to add such corrections while preserving all relevant Ward
identities.  }

{\bf Acknowledgments.}
We acknowledge useful conversations and correspondence with
D. Blaschke, C.D. Roberts, and D. Jarecke.  This work was funded by
the National Science Foundation under Grant Nos.\ INT-9603385,
PHY-9722429 and PHY-0071361 and by the Department of Energy under
Grant Nos.\ DE-FG02-96ER40947 and DE-FG02-97ER41048; it benefited from
the resources of the National Energy Research Scientific Computing
Center, and the Ohio Supercomputer Center.  Both authors are grateful
for the hospitality of the University of Rostock where part of this
work was conducted during several visits and also for the support of
the Erwin Schrodinger Institute for Mathematical Physics, Vienna that
made a visit possible and enjoyable.

%
%
\label{References}

%
%
%
%
%
%
%
\end{document}